\documentclass[12pt,a4paper]{article}
\usepackage{amsmath}
\usepackage{amssymb}
\usepackage{amsfonts}
\usepackage{xcolor}
\usepackage{natbib}
\usepackage[USenglish]{babel} 
\usepackage{setspace}
\usepackage{multicol}
\usepackage{graphicx,caption,rotating}
\usepackage{float}
\usepackage{tabularx}
\usepackage{placeins}
\usepackage{subcaption}
\usepackage{comment}
\usepackage{url}
\usepackage[bookmarks=true,hidelinks]{hyperref}
\expandafter\def\expandafter\normalsize\expandafter{%
    \normalsize
    \setlength\abovedisplayskip{4pt}
    \setlength\belowdisplayskip{4pt}
    \setlength\abovedisplayshortskip{4pt}
    \setlength\belowdisplayshortskip{4pt}
}
\usepackage[]{titlesec}
\titlespacing\section{0pt}{0pt plus 2pt minus 2pt}{0pt plus 2pt minus 2pt}
\titlespacing\subsection{0pt}{0pt plus 2pt minus 2pt}{0pt plus 2pt minus 2pt}
\usepackage[tmargin=2.25cm,bmargin=2.2cm,lmargin=2.2cm,rmargin=2.2cm]{geometry}
\begin{document}
\setcounter{page}{0}
\thispagestyle{empty}
\begin{center}
\vspace{3cm}
\LARGE{Macroeconomic Spillovers of Weather Shocks across U.S. States}\\\vspace{.5cm}
\begin{multicols}{3}
\large{Emanuele Bacchiocchi}\\
\small{University of Bologna}\\
\columnbreak
\large{Andrea Bastianin}\\
\small{University of Milan\\Fondazione Eni Enrico Mattei}\\
\columnbreak
\large{Graziano Moramarco$^{\ast}$}\\
\small{University of Bologna}\\
\end{multicols}
\vspace{2cm}

\end{center}

{\footnotesize{\noindent\textbf{Abstract:} 
		We estimate the short-run effects of weather-related disasters on local economic activity and cross-border spillovers that operate through economic linkages between U.S. states. To this end, we use emergency declarations triggered by natural disasters and estimate their effects using a monthly Global Vector Autoregressive (GVAR) model for U.S. states. Impulse responses highlight the nationwide effects of weather-related disasters that hit individual regions.
		Taking into account economic linkages between states allows capturing much stronger spillovers than those associated with mere spatial proximity. 
		The results underscore the importance of geographic heterogeneity for impact evaluation and the critical role of supply-side propagation mechanisms.
	}\\[.5cm]
	\noindent\textbf{Keywords:} Global VAR, natural disasters, spillovers, weather shocks, United States, climate change \\
	\textbf{JEL Codes:} C32, E32, R11, Q54.}
\vfill
\footnotesize\noindent$^{(\ast)}$\textit{Corresponding author}: Graziano Moramarco, Department of Economics, University of Bologna, Piazza Scaravilli 2, Bologna, Italy. Email: \url{graziano.moramarco@unibo.it}. 
\noindent\textit{Acknowledgments}: 
Emanuele Bacchiocchi gratefully acknowledges financial support from Ministero dell’Università e della Ricerca (MUR), PRIN 20229PFAX5, and the University of Bologna, RFO grants.
Andrea Bastianin acknowledges financial support within the “Fund for Departments of Excellence academic funding” provided by MUR, established by Stability Law, namely “Legge di Stabilita n.232/2016, 2017” - Project of the Department of Economics, Management, and Quantitative Methods, University of Milan.
Graziano Moramarco gratefully acknowledges funding from the European Union and MUR (REACT-EU PON Ricerca e Innovazione 2014-2020 D.M. 1062/2021 CUP J41B21012140007 and GRINS project, code PE0000018 CUP J33C22002910001).
\thispagestyle{empty}
\newpage
\setcounter{footnote}{0}
\newpage
\normalsize
\section{Introduction}\label{sec:intro}
\noindent Climate-related risks are now regarded as major sources of economic vulnerability. These risks are associated with adverse climatic events (\textit{physical risks}) or induced by the process of adjustment towards a low-carbon economy (\textit{transition risks}). Extreme weather events are the main type of physical risk, and at least some of these events have increased in frequency and severity in the U.S.\footnote{The Fifth National Climate Assessment \citep{USGCRP} highlights that extreme cold events have decreased across much of the U.S., while the frequency, intensity, and duration of extreme heat have increased. In addition, other weather events such as heavy precipitation, drought, floods, wildfires, and hurricanes are becoming more frequent and/or severe, with cascading impacts in all parts of the U.S.} \citep{USGCRP} and globally (\citealt{IPCC2023}). Evaluating their economic impacts requires considering not only local effects but also spillovers across interconnected economic systems \citep{BenzieEtAl2021,CashinMohaddesRaissi2017}.

This study analyses the short-run impact of weather-related disasters, or weather shocks for brevity, on state-level economic activity in the United States. Furthermore, the analysis considers the indirect effects of disasters, namely cross-border spillovers resulting from economic interconnections among states and regions. A monthly Global Vector Autoregressive (GVAR) model \citep{pesaranetal04} is employed to capture economic interrelationships between states/regions using bilateral trade flows within the United States. The model incorporates a proxy for weather shocks based on a comprehensive database of emergency declarations triggered by natural disasters. The U.S. represents an interesting case study for assessing the impact of weather-related disasters on the real economy. First, the U.S. territory spans several climate zones, from tropical to polar, and as such it experiences a wide range of extreme weather events, each of which can have a specific impact on the economy. Second, there are a number of reliable indicators that can be taken as proxies for weather shocks, either using geophysical or meteorological datasets, or using damage-based indicators (direct quantification of costs or requests for federal aid by local communities). Third, U.S. data conform to homogeneous standards in terms of how damages are calculated and how aid claims are administered by political authorities. Lastly, the richness of U.S. macroeconomic data makes it possible to study the impact of weather shocks at different levels of geographical disaggregation and to shed light on the transmission channels through which natural disasters affect the real economy.

We find that weather shocks generate direct negative macroeconomic effects at the local level, but also significant negative spillovers to U.S. regions not directly affected. Our results show that weather-related disasters have the potential to act as supply-side shocks, with significant impacts on local economies through disruptions in supply chains and labour markets. The results also indicate that spillovers are stronger when focusing on trade in intermediate goods, thus reinforcing the idea that input linkages across regions and sectors play a key role in amplifying the effects of economic shocks. In addition, we show that geographical disaggregation, parameter heterogeneity and economic linkages between states are key for quantifying spillover effects and the overall impact of weather shocks in the U.S.

We extend the core results along multiple directions to evaluate the robustness and relevance of our findings. We compare our baseline specification with an alternative GVAR model where states are connected by a spatial adjacency matrix, to assess the importance of considering economic interconnections rather than spatial proximity when measuring spillovers. Moreover, since the GVAR model allows for parameter heterogeneity across states, we evaluate the importance of this feature by comparing it with its ``homogeneous-parameter counterpart'', namely a spatial panel dynamic model. Next, we consider a country-wide regression for the U.S. to check whether disaggregation at the state level matters at all. Furthermore, we quantify indirect (second-round) effects on local activity for each state, by comparing responses to weather shocks between the GVAR model -- that allows for spillovers -- and individual state-specific Autoregressive Distributed Lag (ARDL) models where spillovers are implicitly muted. We also address the potential endogeneity of our proxy of weather-related disasters. To this end, we use a two-stage instrumental variable approach that links our indicator to more exogenous, though potentially less precise, proxies for the magnitude of severe weather shocks. These include the number of deaths caused by extreme natural events and a storm indicator that reflects the physical severity of weather phenomena. Lastly, we compare the baseline results with those obtained through the use of two alternative proxies for weather shocks: one based on ex-ante impact estimates generated using the CLIMADA modelling framework \citep{aznar2019climada}, and the other based on the Actuaries Climate Index \citep[see, e.g.,][]{KimMatthesPhan2011}.

The paper is related to several streams of literature. First, it adds to the rapidly growing literature on the macroeconomic effects of extreme weather events and climate change \citep[see, e.g.,][]{BoldinWright2015,BurkeHsiangMiguel2015,ColacitoHoffmannPhan2019,DellJonesOlken2014,ColomboFerrara2024,BCDMO2020WP,FEEM2022_BGVAR,Batten2018,HsiangJina2014}.
In particular, it contributes to the strand of research on cross-border impacts of climate change \citep{BenzieEtAl2021,CarterEtAl2021,feng2024we}. There is ample evidence that the effects of natural disasters propagate through international trade and production networks \citep{BarrotSauvagnat2016,BoehmFlaaenPandalaiNayar2019,CarvalhoNireiSaitoTahbazSalehi2020,feng2024we,ForslidSanctuary2023,Kashiwagi2021}. Our paper is also related to the literature on climate econometrics \citep[see, e.g.,][]{BurkeHsiangMiguel2015,Hsiang2016,KahnEtAl2021}. Focusing on the methodology, it is worth pointing out that the GVAR framework has been used to investigate the international spillovers of El Niño weather shocks by \citet{CashinMohaddesRaissi2017}.

Five papers, to the best of our knowledge, are closely related to our work, in that they focus on the effects of climate and weather extremes on the U.S. economy. 
\citet{MohaddesUSstates2022} and \citet{Natoli23WP} investigate the macroeconomic effects of climate change across the U.S. over the periods 1963–2016 and 1970-2019, respectively. Unlike our analysis, these authors focus on temperature and precipitation anomalies rather than natural disasters, and do not assess spillover effects between states. 
\citet{KimMatthesPhan2011} investigate the effects of severe weather conditions, proxied by the Actuaries Climate Index, on the aggregate U.S. economy over the period 1963-2019. They find that the effects have become significant only in recent years. 
\citet{EickmeierQuastSchüler2024} analyze the impact of natural disasters on the aggregate U.S. economy and financial markets using monthly data from 2000 to 2019, and focusing on extreme temperature events, floods, and storms. Relying on local projections and considering the U.S. economy as a whole, they evaluate the impact of adverse weather shocks, measured as the number of natural disasters in the U.S. in a given month, on a large set of macroeconomic and financial variables. Although their analysis does not consider spillover effects, overall they find a significant and persistent negative impact of natural disasters on the real economy, an increase in the price of food and energy, and a reaction of monetary and fiscal policy aimed at containing the negative macroeconomic impacts. 
\citet{TranWilson2020} mainly focus on the long-run impact of natural disasters on income using a proxy similar to ours, based on U.S. disaster declarations. Relying on panel-data local projections for annual per capita personal income at county level, they obtain little evidence for a negative effect in the short run, but robust evidence of an increase in the longer run, which is coherent with the \lq\lq build back better" scenario. Moreover, checking for geographical spillovers, they find scarce evidence that local aid boosts per capita income in the wider region, in the long run. 
Our paper relies on monthly data and focuses on short-run impacts, thus we complement their analysis based on annual data and concentrating on long-run impacts. Moreover, we study spillover effects based on economic interdependence, while \citet{TranWilson2020} focus on spatial spillovers.

Overall, our findings reinforce the evidence of a significant economic impact of extreme weather conditions, both at local and aggregate level. Although through different methodologies and different levels of aggregation/disaggregation of the data, our results are in line with other contributions in the literature, like \cite{EickmeierQuastSchüler2024}, \cite{ColacitoHoffmannPhan2019}, \cite{Natoli23WP}, \cite{MohaddesUSstates2022}, \cite{KimMatthesPhan2011}; this latter finds negative effects mainly for the last part of the sample, which approximately corresponds to our period of investigation. A notable exception is represented by \cite{TranWilson2020}, who instead find slightly negative but statistically insignificant short-run (contemporaneous) responses of real activity indicators to adverse climate shocks.

The rest of the paper is organized as follows. 
Section \ref{sec:data} presents the data, with a particular focus on our proxy for severe weather shocks; 
Section \ref{sec:model} discusses the econometric model; 
Section \ref{sec:results} presents the main results;
Section \ref{sec:moreres} provides further results and robustness checks;  
Section \ref{sec:conclusions} concludes. 
An online Appendix with additional results completes the paper.

\section{Data}\label{sec:data}

\subsection{Weather-related disasters in the U.S.}\label{subsec:data_weather}

\noindent We build a novel monthly proxy of weather shocks using the Disaster Declarations Summary dataset maintained by the U.S. Federal Emergency Management Agency (FEMA). FEMA, established in 1979 with an executive order signed by President Carter, is part of the Department of Homeland Security and is responsible for coordinating the federal government's relief efforts after natural or man-made domestic disasters. All federally declared disasters since 1953 are collected in the Disaster Declarations Summary. For each declaration, the database provides detailed information on the states and counties affected by the disaster, its beginning and end dates, as well as the type of disaster.

Because the Congress has broadened FEMA's mandate over the years, the number of declarations has also grown since its inception and in particular after 1988. In fact, the Stafford Disaster Relief and Emergency Assistance Act of 1988 established that a presidential declaration triggers financial and physical assistance through FEMA. To avoid issues related to this policy shift induced by the Stafford Act, we restrict the sample to events from January 1990 through December 2019. We consider three types of disaster declarations that differ in terms of the events triggering them, their scope, as well as the amount of funds and type of assistance provided. These are: Emergency Declarations, Major Disaster Declarations, and Fire Management Assistance Declarations.\footnote{The dataset is available online at: \href{https://www.fema.gov/openfema-data-page/disaster-declarations-summaries-v2}{https://www.fema.gov}. For details about the emergency declaration process, see: \href{https://www.fema.gov/disaster-declaration-process}{https://www.fema.gov/disaster-declaration-process}.} 

\begin{figure}
    \caption{Emergency declarations by state and type of event: 1990-2019}
    \centering
    \includegraphics[width=.33\linewidth]{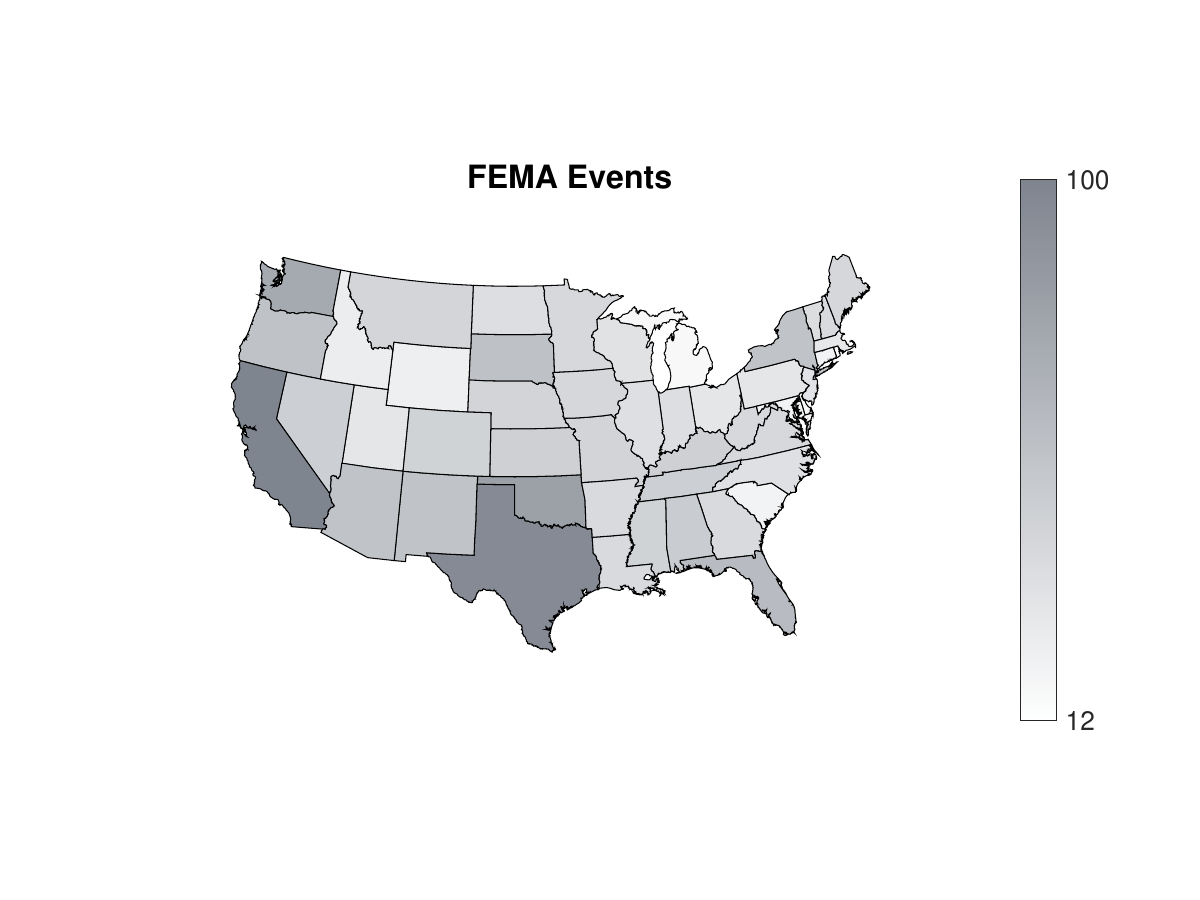}%
    \includegraphics[width=.33\linewidth]{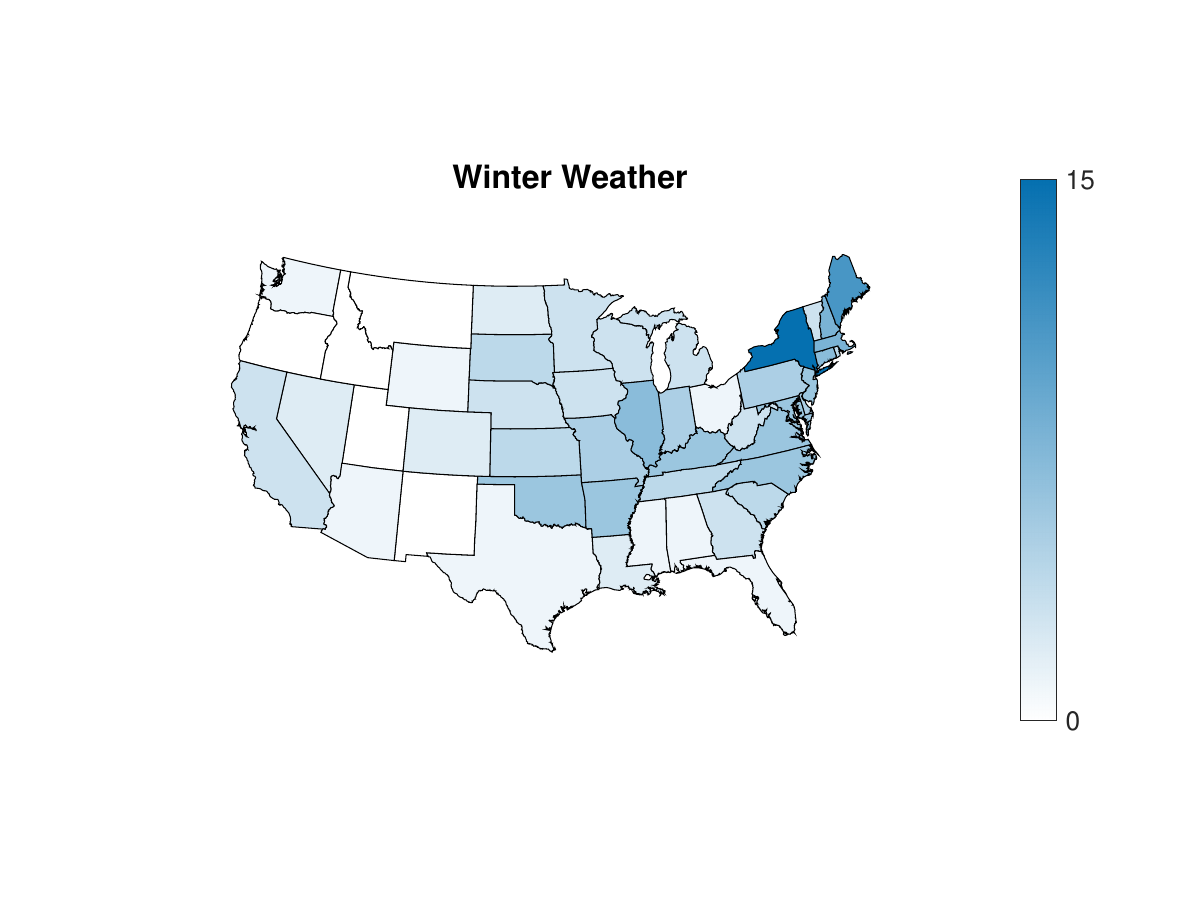}%
        \includegraphics[width=.33\linewidth]{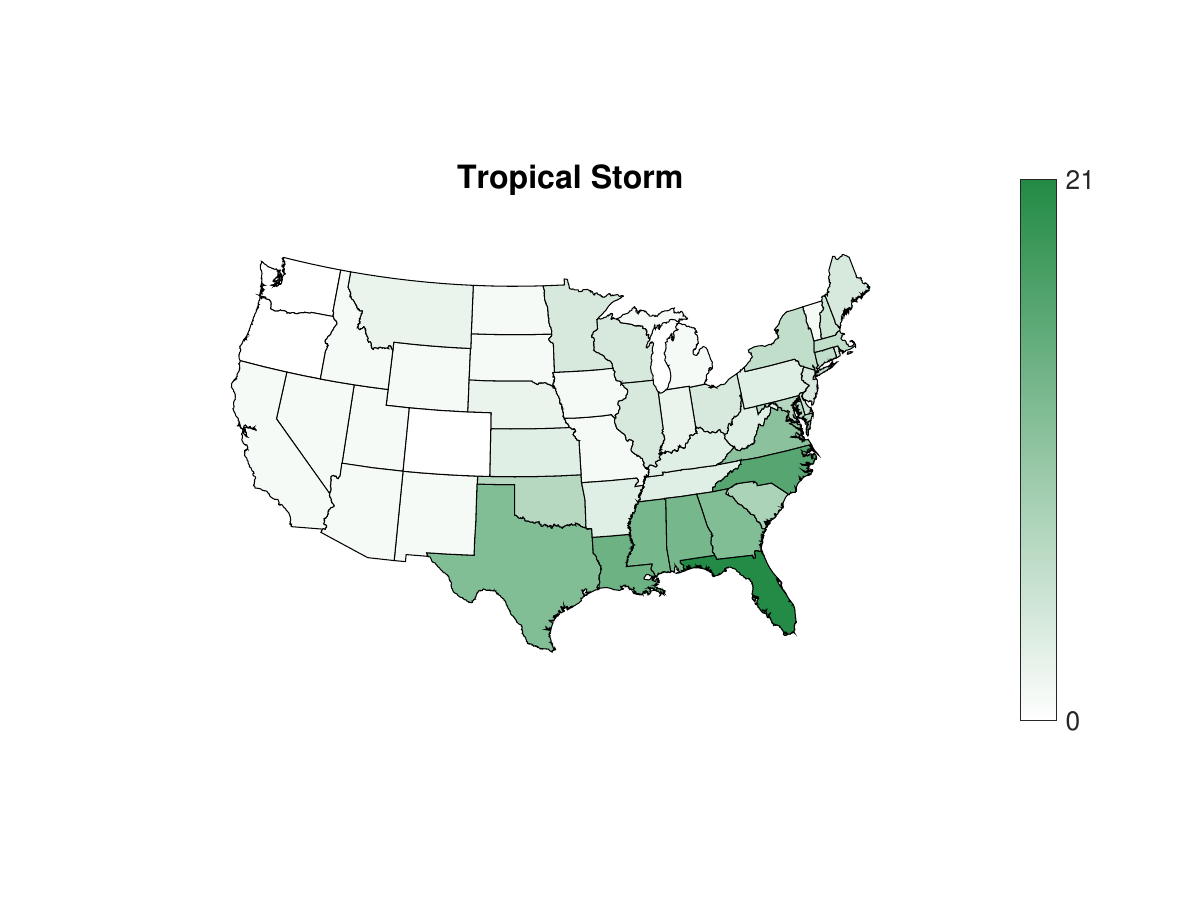}\\
    \includegraphics[width=.33\linewidth]{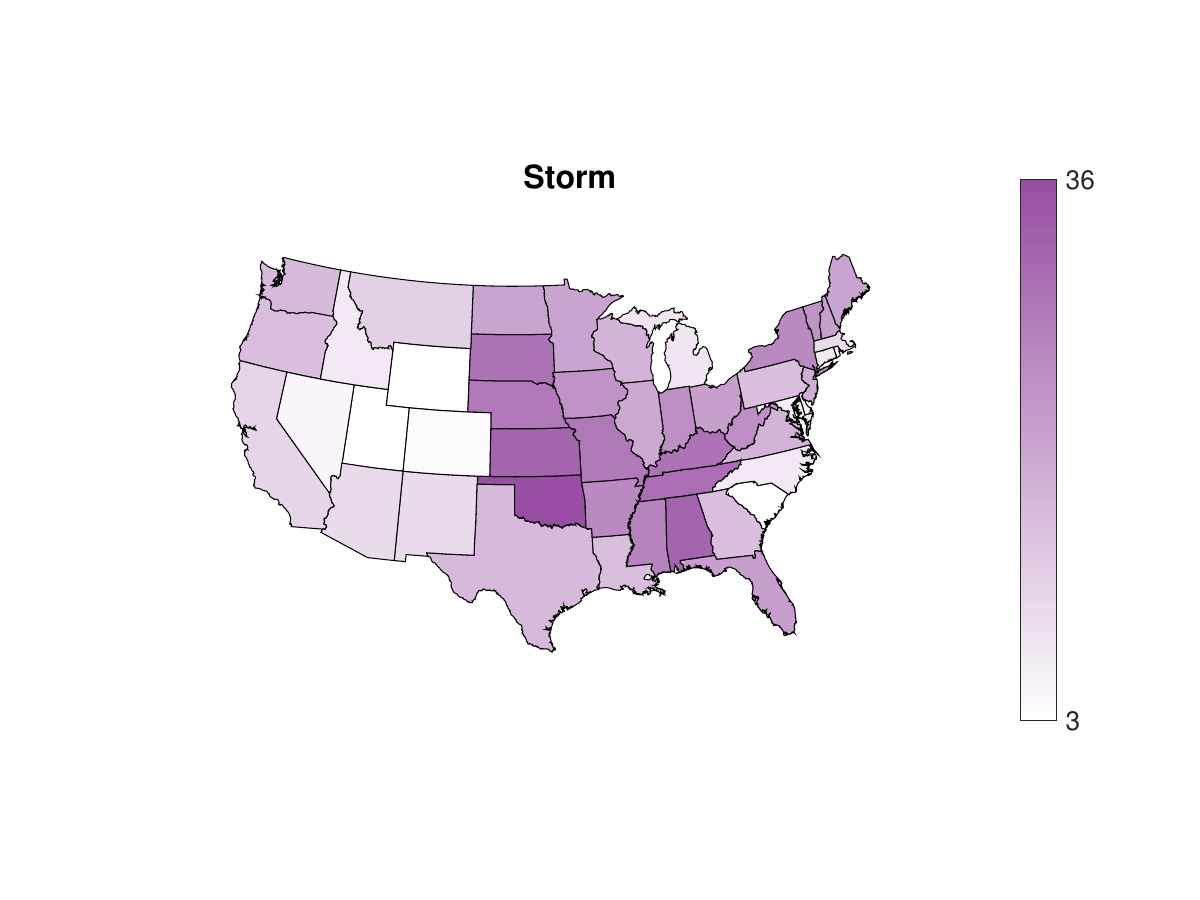}%
        \includegraphics[width=.33\linewidth]{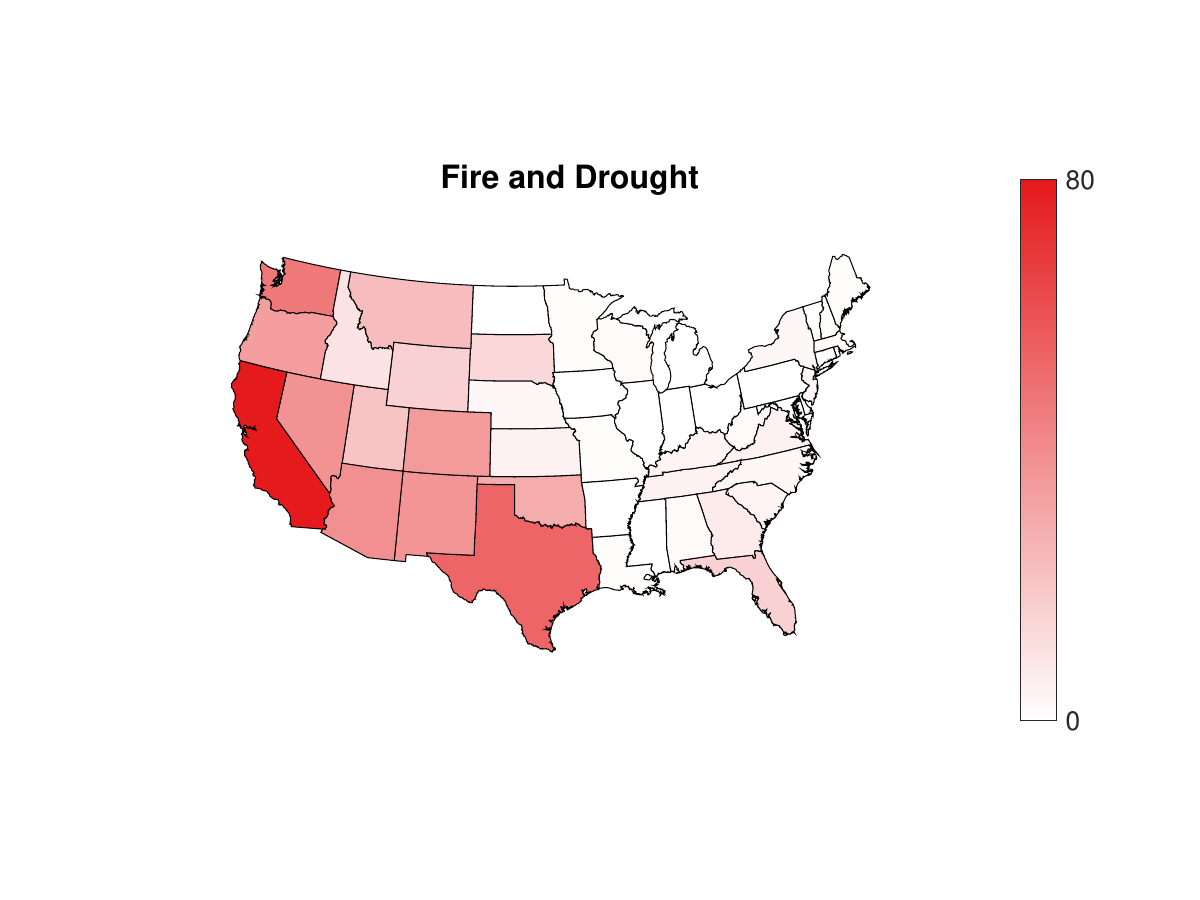}%
    \includegraphics[width=.33\linewidth]{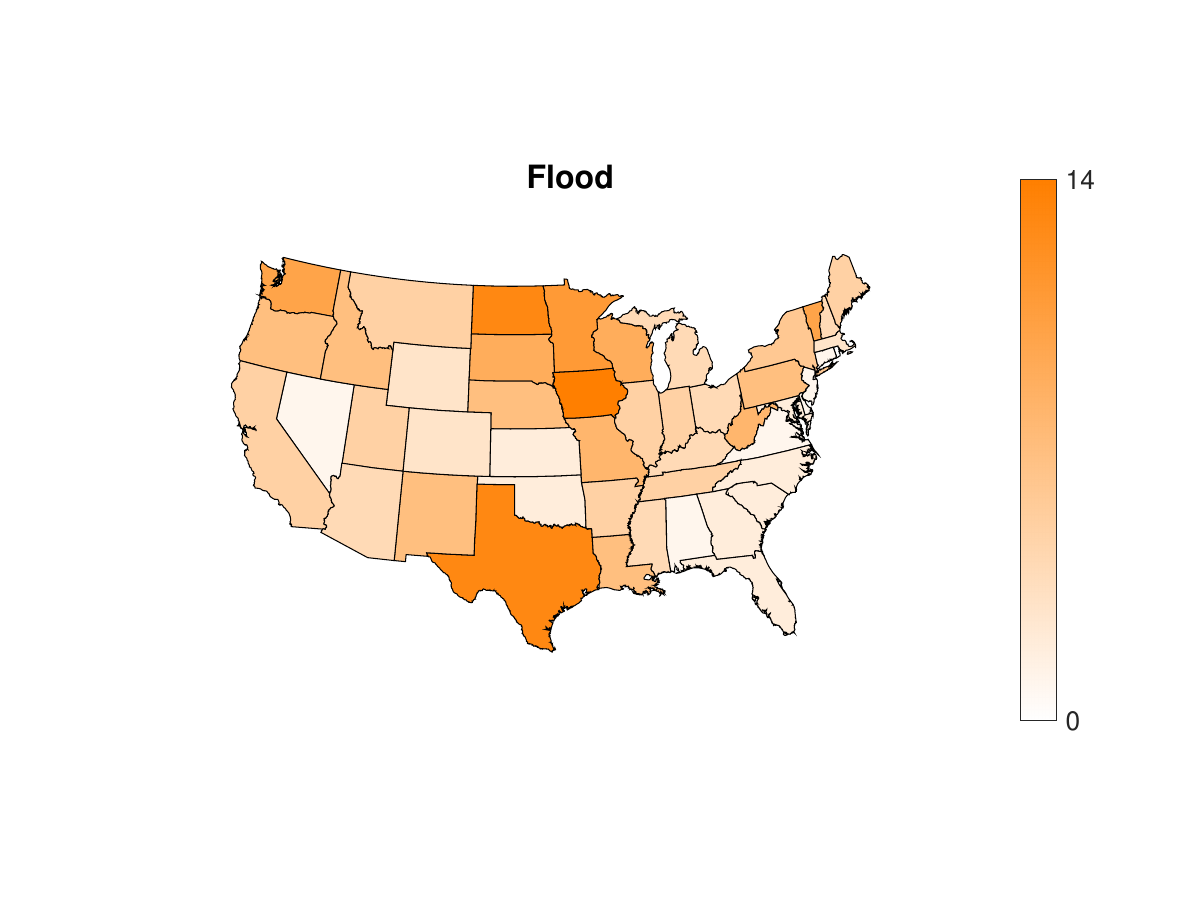}\\
    \label{fig:mapstates}
    \caption*{\scriptsize{\textit{Notes}: The maps show the count of emergency declarations (total and by type of event). Colors range from lighter to darker as the count of events per state grows. Minimum and maximum values are shown in the color bar. Alaska and Hawaii are omitted to improve visual clarity.}}
\end{figure}
\FloatBarrier

We focus on declarations caused by events that are directly related to weather conditions. In particular, we consider the following main types of events that group the weather-related declarations in FEMA's database: winter weather (i.e., freezing, ice storms, snow), tropical storms (i.e., hurricanes, typhoons and tornadoes), storms (i.e., coastal and severe storms), fires, floods, and droughts. Most events show the expected seasonal patterns, with storms and fires being the most recurrent types of event. Figure \ref{fig:mapstates} shows the count of declarations by state and by type. Darker colors indicate a higher number of events. If we do not distinguish by event type -- upper left panel -- we can see that emergency declarations are quite evenly distributed across states. The same holds true for storms and floods. On the contrary, declarations triggered by severe winter weather, tropical storms, fires and drought are geographically concentrated in the northeast, southeast, and west of the U.S., respectively.\footnote{Further descriptive statistics on weather events and alternative representations of the maps in Figure \ref{fig:mapstates}, either based on the count of declarations weighted by state population density or aggregated over U.S. climate regions, are provided in the online Appendix.}

\bigskip

\begin{figure}
    \centering
        \caption{Weather shocks for the U.S.}
    \includegraphics[width=\textwidth]{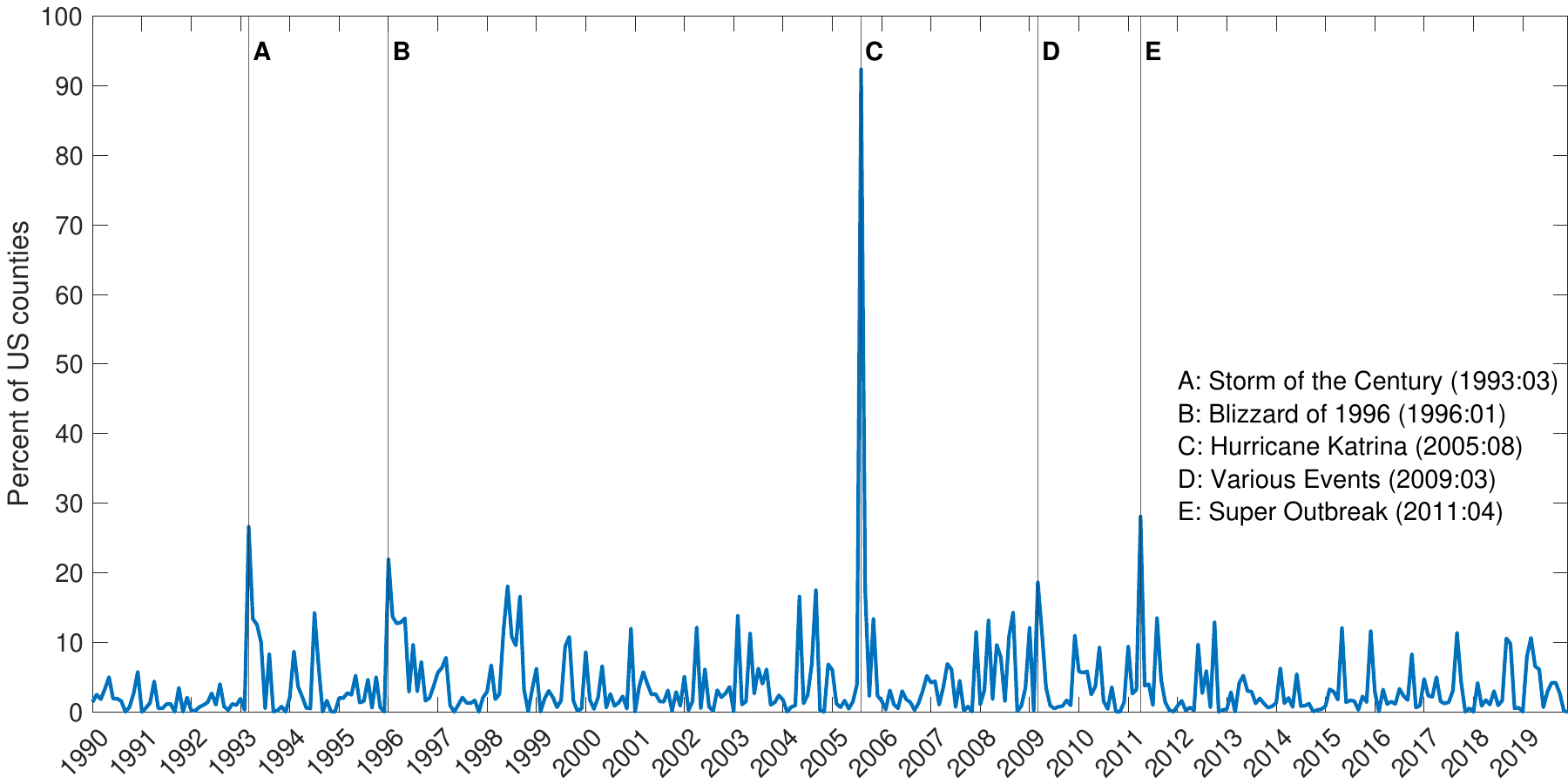}
    \caption*{\scriptsize\textit{Notes}: For Hurricane Katrina, emergency declarations were issued in almost all states. In states not directly hit by the hurricane, the declarations go under the header ``Hurricane Katrina Evacuation''.}
    \label{fig:usshocks}
\end{figure}

\noindent{\textit{Measuring weather shocks with disaster declarations.}} We construct our proxy of weather shocks as follows. First, for each state $i$ we create a dummy variable that takes value one if a weather-related disaster was declared in month $t$, and zero otherwise ($emergency_{it}$). Second, we multiply this dummy variable by a weight, given by the ratio between the number of counties affected by the disaster ($hit_{it}$) and the total number of counties (or county equivalents) in state $i$ ($counties_i$):
\begin{equation}
 s_{it} =emergency_{it}\times \frac{hit_{it}}{counties_i}   
\end{equation}
Therefore, our state-level weather shock has unit value only if, in a given month, emergency declarations are issued for all the counties in a state. To calculate an aggregate weather shock for the U.S., we consider the number of counties hit as a share of the total number of counties in the 50 states considered in the analysis ($N_c = 3142$):\footnote{The list of U.S. counties and county equivalents is sourced from: \href{https://en.wikipedia.org/wiki/County\_(United\_States)}{https://en.wikipedia.org/wiki/County}.}
\begin{equation}
 s_{t} = \frac{1}{N_c}\sum_{i=1}^{50} emergency_{it}\times hit_{it}   
\end{equation}
The weather shock series for the U.S. is displayed in Figure \ref{fig:usshocks}, where we highlight the five largest shocks over the period 1990-2019. Hurricane Katrina is by far the largest event, as measured by the share of counties where emergency declarations were issued. Figure \ref{fig:usshocks} also shows that the number of declarations for the U.S. is highly volatile, but there is no visible trend, nor structural breaks in the series. The weighting of events based on the percentage of affected counties in the state is intended to highlight the intensity of disasters rather than the importance of individual counties in the economy of a given state.\footnote{In the online Appendix, we consider alternative weighting schemes and proxies of weather shocks. See also Section \ref{sec:aci}.}

To sum up, the FEMA dataset includes all disasters ``\textit{of such severity and magnitude that effective response is beyond the capabilities of the state 
$(\ldots)$ and that supplemental federal assistance is necessary}.''\footnote{Source: \url{ https://www.fema.gov/disaster/how-declared}} Extreme weather events measured in this way represent ``proper'' shocks in the parlance of \citet{Ramey2016}: (\textit{i}) they are exogenous with respect to the current and lagged endogenous variables of the model; (\textit{ii}) they are uncorrelated with other exogenous shocks (e.g., technological, monetary or fiscal policy shocks); and (\textit{iii}) they are generally unanticipated. 

\subsection{State-level economic variables}\label{subsec:data_economy}
\noindent\textit{State-level economic activity.} Our aim is to capture the impact of weather-related disasters on broad measures of economic activity that allow assessing the presence of spillover effects. Unfortunately, data on Gross State Product is only available at a quarterly frequency from 2005, resulting in a relatively small number of available observations. Moreover, working with monthly data appears preferable in this context, since the duration of extreme weather events is generally shorter than one month, and so using lower-frequency data may lead to temporal aggregation bias.
The proxy of economic activity developed by \citet{BaumeisterLeivaLeonSims2022}, referred to as weekly Economic Conditions Indicator (ECI), is suitable for our purposes. This is a composite indicator based on mixed-frequency dynamic factor models covering multiple dimensions of state-level economic activity (mobility, labor market, real activity, expectations measures, financial and households indices). Since the ECI is scaled to match four-quarter growth rates of U.S. real GDP, and we are interested in measuring the economic effects of weather shocks at a monthly frequency, we take the monthly average of the weekly index and label it Monthly Economic Conditions Indicator (MECI) hereafter. Unlike other indicators of economic activity available at the state level on a monthly basis - such as employment or hours worked - the MECI includes a broad set of both standard and non-standard variables (e.g., electricity consumption, oil rig counts, vehicle miles traveled) that allow for a more comprehensive measure, thereby enabling a more thorough analysis of spillover effects. See \citet{bokun2023fred} for an overview of state-specific variables for the U.S.\footnote{In the online Appendix, we rely on more disaggregated indicators of economic activity (e.g., the number of hours worked in manufacturing) as alternative dependent variables in the GVAR, in place of MECI. The additional results corroborate the baseline ones.
However, as expected, using less comprehensive measures of economic activity leads to estimate weaker spillover effects.}

Note that we are interested in studying the effects of extreme weather, therefore it is important to assess the presence of seasonality in the MECI. The individual components of the ECI are seasonally adjusted when appropriate \citep[please refer to][for details on seasonal adjustment]{BaumeisterLeivaLeonSims2022}. Nevertheless, we further check that there are no seasonal effects left in the monthly version of the index by estimating state-specific autoregressive models augmented with month-of-the-year dummy variables. The $F$-tests indicate that for all states, but North Carolina, month-of-the-year dummies are jointly non-significant, and therefore seasonal factors do not affect our results.

\bigskip

\noindent\textit{Interstate trade flows.} In GVAR models, economic interrelationships between geographic entities are most commonly measured using trade flows. To proxy bilateral trade flows between U.S. states, we use data from the latest release of the Commodity Flow Survey (CFS) in 2017. The CFS is jointly elaborated by the Bureau of Transportation Statistics (U.S. Department of Transportation) and the U.S. Census Bureau (U.S. Department of Commerce) every five years as part of the Economic Census. This survey is the primary source of data on domestic freight shipments by establishments in mining, manufacturing, wholesale, and selected retail and services trade industries (namely, electronic shopping and mail-order houses, fuel dealers, publishers) located in the U.S. We use these data to construct weight matrices that summarize network links between states. In particular, to calculate the weight of state $j$ in the network of state $i$, we use the sum of bilateral imports and exports between the two states (i.e., amounts of CFS shipments of goods from $j$ to $i$ and from $i$ to $j$, respectively), as a ratio of total trade of state $i$ (imports into $i$ plus exports from $i$) with all other states in the U.S.

\begin{figure}[t]
    \centering
    \caption{NOAA U.S. Climate Regions and network visualization of trade flows between states}
        \includegraphics[width=.495\textwidth]{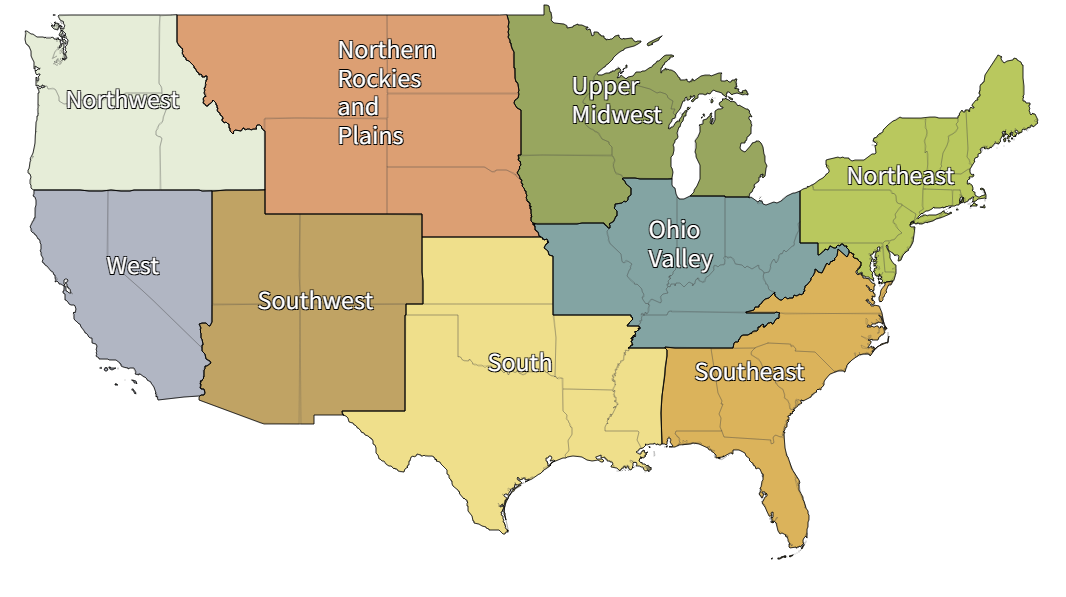}
        \includegraphics[width=.495\textwidth]{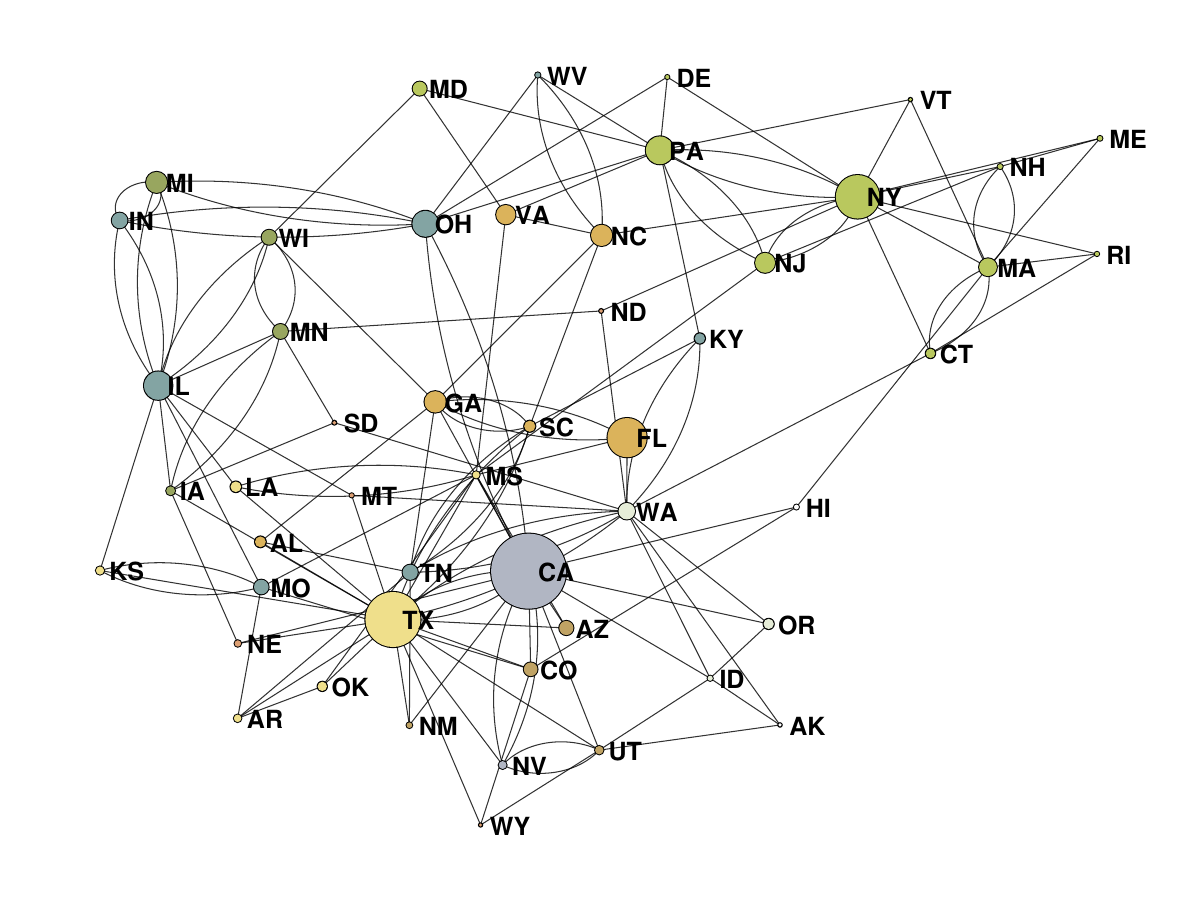}
        \caption*{\scriptsize\textit{Notes:} The left panel shows the classification of U.S. states into nine climate regions, provided by the National Oceanic and Atmospheric Administration (source: \url{https://www.ncei.noaa.gov/access/monitoring/reference-maps/us-climate-regions}). Alaska and Hawaii are not considered in this classification but are kept in the analysis. The right panel provides a network visualization of major bilateral trade flows between U.S. states.}
        \label{fig:NOAA_regions}
\end{figure}

When reporting the results, we will aggregate states into climate regions. To this end, we will use the classification adopted by the National Oceanic and Atmospheric Administration (NOAA). Specifically, scientists at the NOAA National Centers for Environmental Information (NCEI) have identified nine climatically consistent regions \citep{usclimatereg}. On the left panel of Figure \ref{fig:NOAA_regions} we show these nine climate regions, which are: Northeast (NE), Southeast (SE), South (S), Upper Midwest (UMW), Ohio Valley (OV), Northern Rockies and Plains (NP), Southwest (SW), West (W), and Northwest (NW).

A network representation of the interstate trade flows, instead, is provided in the right panel of Figure \ref{fig:NOAA_regions}. Visual clutter is reduced using CFS data to determine, for each state, the three most important trade partners and representing the network as an undirected graph. Nodes are proportional to the size of the local economy, as proxied by state employment in 2017 sourced from the FRED-SD database \citep{bokun2023fred}, and have the same color as the climate region to which they belong, as shown in the left panel of the figure. The network reflects the underlying spatial proximity of states as well as their centrality to the U.S. economy as a whole. For instance, we can observe a cluster of northeastern economies tightly interconnected on the right portion of the graph. Moreover, it is evident that New York, Texas, Florida and California not only represent the largest economies in terms of employment, but are also heavily connected with several other states, thus playing a pivotal role in the entire U.S. economy.

\section{A GVAR model for the U.S. economy}\label{sec:model}

\noindent We construct a GVAR model for the U.S. economy by aggregating state-specific autoregressive models (ARX*) that include rest-of-U.S. (RoUS) exogenous variables, along with weather shocks. RoUS aggregate variables are calculated as weighted averages across all other states, using inter-state trade shares as weights. Since RoUS aggregates are treated as weakly exogenous in each ARX* model, estimation can be performed at the state level, thus avoiding overparametrization problems. Let $y_{it}$ denote an indicator of economic activity for state $i$ at time $t$, and $s_{it}$ be our state-level weather-shock proxy. For ease of notation, in the following discussion we consider a model with one lag. The ARX* for state $i$ is then given by:
\begin{equation}\label{eq:ARX}
    y_{it} = \alpha_{i} +  \beta_{i} y_{i,t-1} + \gamma_{0i} y_{it}^\ast + \gamma_{1i} y_{i,t-1}^\ast + \theta_i s_{it} + u_{it} 
\end{equation} 
where $y_{it}^\ast$ denotes RoUS average economic activity, 
$\alpha_{i}$, $\beta_{i}$, $\gamma_{0i}$, $\gamma_{1i}$ and $\theta_{i}$ are parameters to be estimated, and $u_{it}$ is an i.i.d. error with mean 0 and variance $\sigma_i^2$. As shown by \citet{DeesDiMauroPesaranSmith2007}, the GVAR can be derived as an approximation to a global unobserved common factor model, where common factors are captured by cross-country averages. In our model, cross-state averages capture factors that affect the U.S. economy as a whole, such as international demand for U.S. goods.

Let $\mathbf{W}_i$ denote the weight matrix for state $i$, such that:
    \small
    \begin{equation}\label{eq:GVAR_weights}
        \left(\begin{array}{c}
            y_{it}\\
            y_{it}^\ast 
        \end{array}\right)
        = 
        \mathbf{W}_i 
        \Big(y_{1t} \hspace{0.3cm} y_{2t}\hspace{0.3cm} \cdots \hspace{0.3cm} y_{Nt}\Big)^\prime
        = \mathbf{W}_i \mathbf{y}_t
    \end{equation} \normalsize
\hspace{-25pt}
where $N=50$ is the number of states and $\mathbf{y}_t$ is the country-wide vector of endogenous variables. Specifically, $\mathbf{W}_i$ contains the weights of all states in the trade network of state $i$, measured using CFS data introduced in Section \ref{sec:data}. As mentioned before, the weight of state $j$ in the network of state $i$ is calculated as the sum of bilateral imports and exports between $i$ and $j$ as a ratio of total trade of state $i$ with all other states in the U.S.

Stacking all ARX* models, we get a country-wide model:
\vspace{0pt}
\begin{equation}
    \mathbf{G} \mathbf{y}_t = \boldsymbol{\alpha} + \mathbf{H} \mathbf{y}_{t-1} + \boldsymbol{\Theta} \mathbf{s}_t + \mathbf{u}_{t}
\end{equation}  		
where
\footnotesize     	  \begin{equation}
    \hspace{-10pt}  
    \mathbf{G} = \begin{pmatrix} \left(1, -\gamma_{01} \right) \mathbf{W}_1 \\  \left(1, -\gamma_{02} \right) \mathbf{W}_2 \\  \vdots \\  \left(1, -\gamma_{0N} \right) \mathbf{W}_N \end{pmatrix}\hspace{35pt}
    \mathbf{H} = \begin{pmatrix} \left(\beta_{1}, \gamma_{11} \right) \mathbf{W}_1 \\  \left(\beta_{2}, \gamma_{12} \right) \mathbf{W}_2 \\ \vdots \\  \left(\beta_{N}, \gamma_{1N} \right) \mathbf{W}_N \end{pmatrix} \hspace{35pt}
    \boldsymbol{\Theta} = \begin{pmatrix} \theta_1 & 0 & \dots & 0 \\  0 & \theta_2 & \dots  & 0  \\ \vdots &  \vdots & \ddots & \vdots \\  0  & 0  & \dots & \theta_N \end{pmatrix}
\end{equation} \normalsize
\vspace{5pt}
\hspace{-20pt}
while $\boldsymbol{\alpha}$, $\mathbf{s}_t$ and $\mathbf{u}_{t}$ are vectors of stacked constants, weather shocks and errors, respectively. By inverting the matrix $\mathbf{G}$, we obtain a reduced-form VAR with exogenous variables (VARX) for the entire U.S. economy:
\begin{equation}
    \mathbf{y}_t = \mathbf{c} + \mathbf{F} \mathbf{y}_{t-1} + \boldsymbol{\Lambda} \mathbf{s}_t + \boldsymbol{\varepsilon}_t
\end{equation}
where $\mathbf{c} = \mathbf{G}^{-1} \boldsymbol{\alpha}$, 
$\mathbf{F} = \mathbf{G}^{-1} \mathbf{H}$, 
$\boldsymbol{\Lambda}  = \mathbf{G}^{-1} \boldsymbol{\Theta}$,
and $\boldsymbol{\varepsilon}_t = \mathbf{G}^{-1} \mathbf{u}_{t}$.
It is important to remark that matrix $\boldsymbol{\Lambda}$ is not diagonal: its off-diagonal elements capture \textit{contemporaneous} cross-border effects of weather shocks.

\section{Main results}\label{sec:results}

\noindent We estimate our monthly GVAR on data from January 1990 to December 2019. We exclude the subsequent period to prevent our results from being biased by the COVID-19 pandemic. 
We derive our baseline results using the MECI as a measure of local economic activity. Unit root tests indicate that the level of the MECI is non-stationary or near unit root for a number of states, and stationary for others.
Therefore, we transform the MECI indices by taking their first differences and estimate each ARX* model \eqref{eq:ARX} by OLS.\footnote{Since not all MECI indices are non-stationary, we cannot represent our U.S. GVAR model as a cointegration model as in \citet{pesaranetal04}. 
Moreover, we have tried estimating the GVAR on the levels of the MECI, treating them as stationary, but the model turns out to have explosive roots in this case. Reliance on first differences of MECI implies that our model does not have an error correction form. For the same reason, we cannot rely on the usual likelihood ratio (LR) test for weak exogeneity of $y_{it}^{*}$ used in the cointegrated GVAR literature, which tests the significance of loading coefficients in error-correction country-specific models (see, e.g., \citealt{DeesDiMauroPesaranSmith2007}). However, we perform Granger causality tests on $y_{it}$ and  $y_{it}^{*}$ in each state, separately. The tests indicate that, for the vast majority of states, including the ones with the largest economies, the domestic variable $y_{it}$ does not Granger cause the rest-of-U.S. variable $y_{it}^{*}$, while $y_{it}^{*}$ Granger causes $y_{it}$, as expected. Details in the online Appendix.}
We rely on a GVAR model with 2 lags of first-differenced MECI, as suggested by minimization of the Bayesian information criterion, which selects at most 2 lags for almost all states.\footnote{For robustness, we have also estimated the model using, alternatively, lag orders of 1 and 3. The results remain very similar in both cases. Furthermore, in consideration of the monthly frequency of the data, we have also estimated the model with 12 lags. The results are qualitatively similar also in this case.}

We assess the direct and indirect macroeconomic effects of weather shocks by means of impulse response functions (IRFs) estimated on a sample that runs from April 1990 through December 2019, for a total of 357 observations (i.e., after taking first differences of MECI and considering two lags). While we carry out the impulse-response analysis at the state level, in what follows, to summarize results, we aggregate the estimated individual IRFs into those for nine U.S. climate regions, as previously shown in Figure \ref{fig:NOAA_regions}. Note that while Alaska and Hawaii are not considered in this classification, we do retain them in the GVAR model.

The impulse response functions are obtained by simulation, i.e., by projecting \textit{h} steps ahead the fitted indicator of economic activity, with and without the impulse of the weather shock, and taking the difference between the two projected paths. When simulating a weather shock, we assume that all states in a region are contemporaneously hit, and all counties within each state. As a matter of fact, Figure \ref{fig:usshocks} highlights that major weather events such as the Storm of the Century in 1993, the Blizzard of 1996, Hurricane Katrina in 2005 and the Super Outbreak in 2011 have triggered emergency declarations in a very large portion of the U.S. This is thus an extremely adverse but realistic scenario, that allows us to interpret the results as ``upper bounds'' (in absolute value) of cross-regional spillover effects.

\subsection{The impacts of weather shocks on economic activity}\label{sec:mainresults}

\begin{figure}[t]
        \caption{Impulse response functions - Monthly Economic Conditions Indicators}
              \centering
              \includegraphics[width=\textwidth]{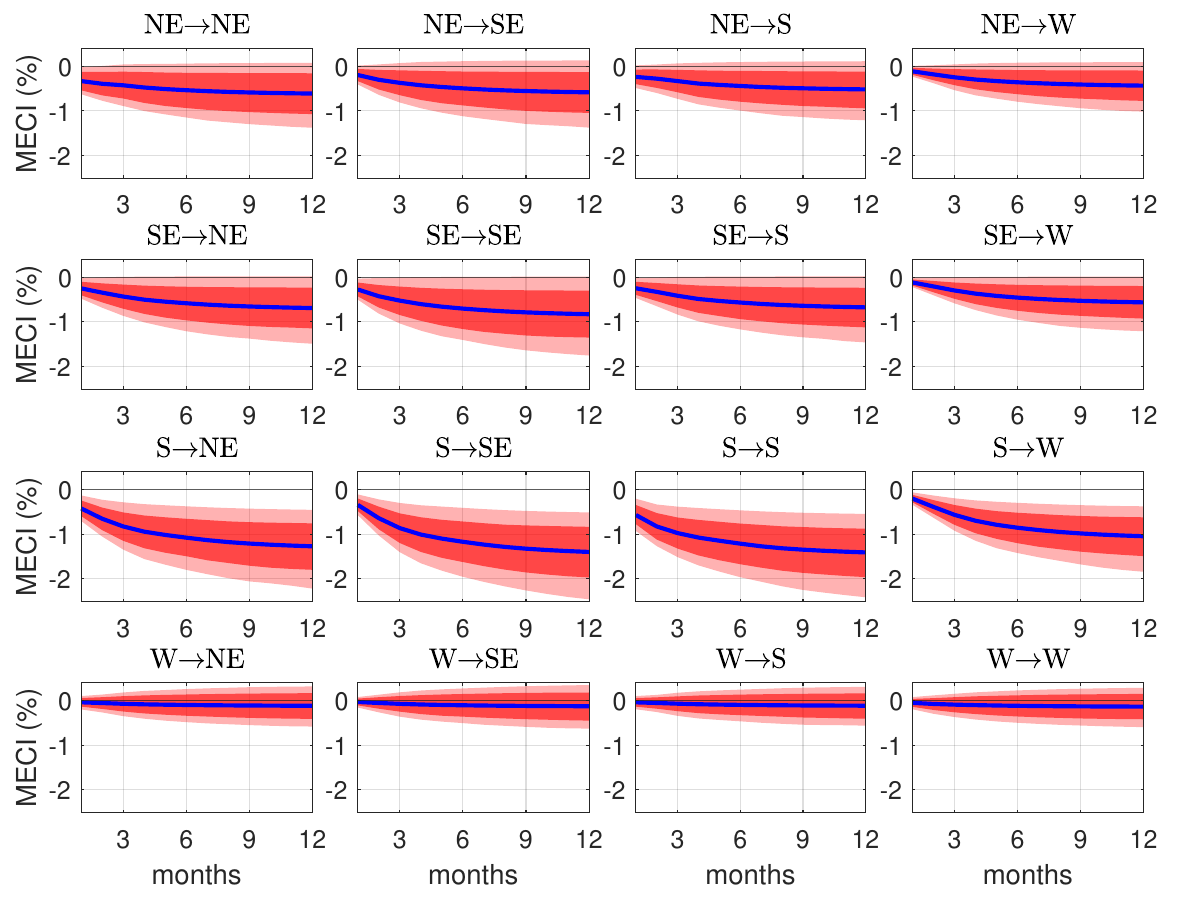}
	            \caption*{\scriptsize\textit{Notes:} Responses of the Monthly Economic Conditions Indicator (MECI) to a weather shock hitting 100\% of counties in the shocked region. The horizontal axis measures months, the vertical axis the responses in percentage points. Blue solid lines indicate bootstrap means and light-red (dark-red) areas the corresponding 2.5th and 97.5th (10th and 90th) percentiles. Four climate regions are considered: Northeast (NE), Southeast (SE), South (S), and West (W). In the titles, the first region indicates where the shock originates, while the second one is where it propagates.}
            \label{fig:irf}
	\end{figure}

\begin{figure}[t]
\caption{IRFs of Monthly Economic Conditions Indicators, impact on month of the shock}
\centering
    \includegraphics[width=1\linewidth]{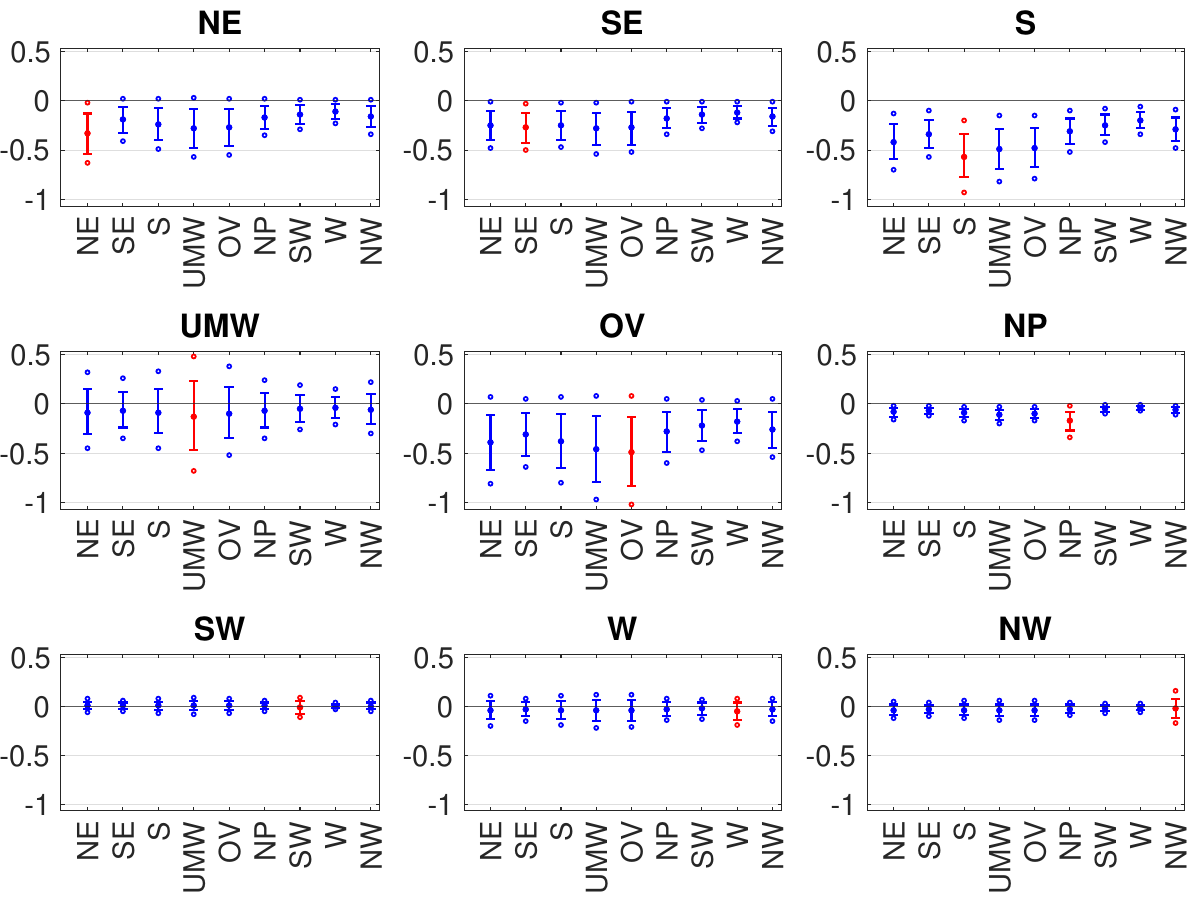}
\caption*{\scriptsize\textit{Notes:} The figure displays IRF bootstrap means, along with the corresponding 2.5th and 97.5th (10th and 90th) percentiles, represented as points (bars). The horizontal axis specifies the regions responding to shocks, while the title identifies the origin of the shocks.}
    \label{fig:figirf_all_impact}
\end{figure}

\begin{figure}[t]
\caption{IRFs of Monthly Economic Conditions Indicators at 12-month horizon}
\centering
    \includegraphics[width=1\linewidth]{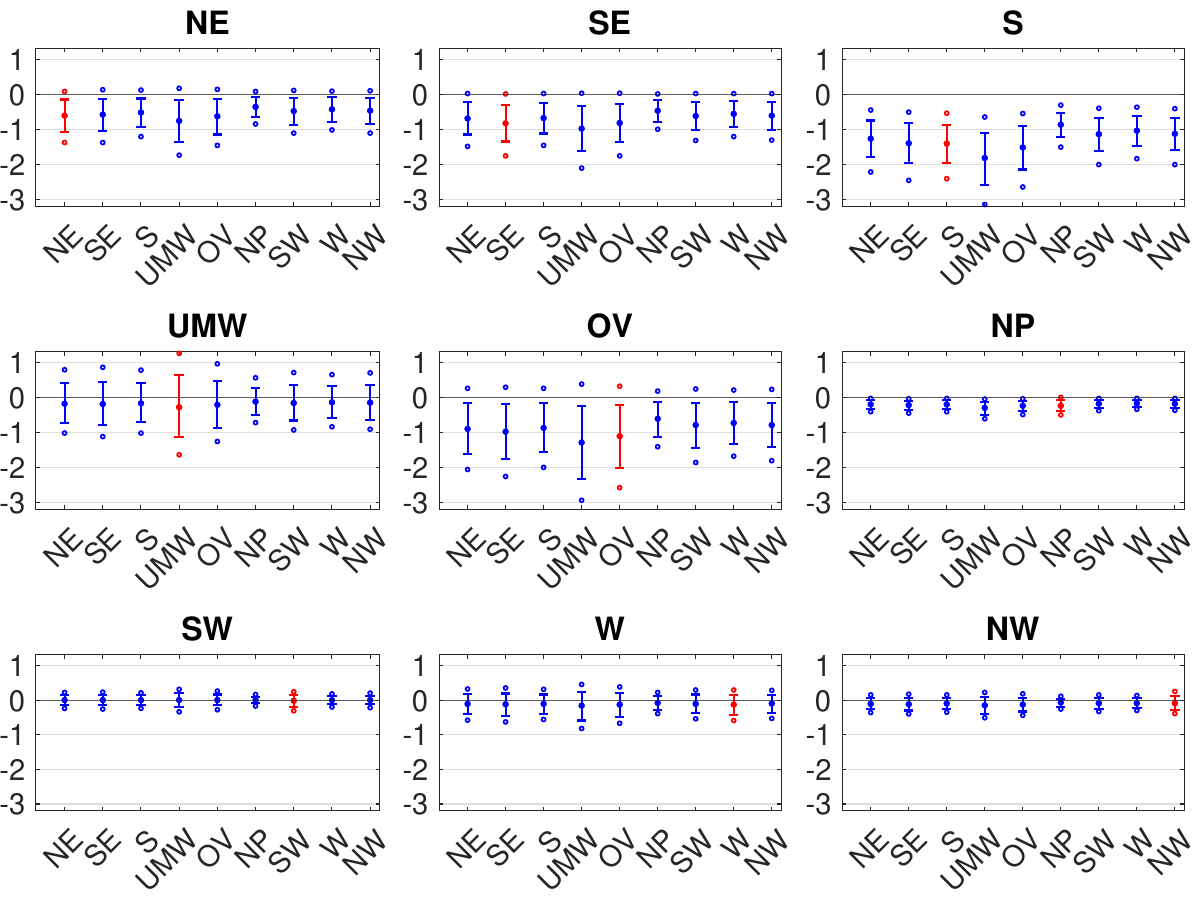}
\caption*{\scriptsize\textit{Notes:} See notes to Figure \ref{fig:figirf_all_impact}.}
    \label{fig:figirf_all}
\end{figure}

\noindent We assess the responses of MECI to a region-wide weather shock in Figures \ref{fig:irf}-\ref{fig:figirf_all}. While the GVAR model is estimated using the first differences of MECI, we cumulate the IRFs to obtain the responses in terms of level of the index, which is expressed as a year-on-year percent growth rate (see section \ref{subsec:data_economy}). Thus, an IRF value of -1 means that a shock reduces economic activity by about 1\% compared to one year before. As discussed in the Appendix, where we also report the IRFs for the first differences of the index ($\Delta \text{MECI}$), the impact on month-on-month growth rates generally vanishes within a year after a shock. Figure \ref{fig:irf} shows the IRFs for a subset of regions (namely, NE, SE, S, and W), while Figure \ref{fig:figirf_all_impact} reports the IRFs for all regions in the month when the shock occurs, and Figure \ref{fig:figirf_all} reports the IRFs for all regions one year after the shock, which seems a reasonable time frame to assess the magnitude of macroeconomic effects induced by weather disasters. We compute mean IRFs and include both the 80\% (i.e., based on the 10th and 90th percentiles) and 95\% (i.e., based on the 2.5th and 97.5th percentiles) confidence intervals using a bootstrap procedure for GVAR models \citep[see][for details]{DeesDiMauroPesaranSmith2007,DeesHollyPesaranSmith2007}.\footnote{The bootstrap procedure is non-parametric, so we do not need to make any assumption on the distribution of errors $u_{it}$ in equation \eqref{eq:ARX}. In the Appendix, we also report results for the 5th and the 95th percentiles (i.e., the 90\% confidence interval).} 
Since the results are qualitatively comparable, in what follows we comment on the results focusing on the 80\% confidence interval.

Remarkably, the responses of economic activity are generally negative, both within the shocked region and in regions not directly affected by the weather shock. Except for Western regions (W, NW, and SW), all regions generate significant spillovers. 
The cumulative IRFs converge within approximately one year after the shock, which is consistent with monthly economic activity being affected for periods lasting up to one year.

 \FloatBarrier

As already said, Figure \ref{fig:figirf_all_impact} and Figure \ref{fig:figirf_all} report the mean bootstrap responses along with confidence intervals on the month of the shock and one year after the shock, respectively. The largest spillovers (in absolute values) are generated by the South (S), Southeast (SE) and Ohio Valley (OV) regions. Concerning the magnitude of the effects, the mean responses after one year are generally in the range from -1.8 to 0. This indicates, approximately, a drop of up to 1.8\% in year-on-year growth rates of economic activity, in response to the region-wide weather shock. 
In interpreting the results, it should be kept in mind that we are reporting responses to an upper-bound shock hitting 100\% of counties in all states within a region. Of course, extreme events typically hit only a fraction of counties. 
On average, in months when weather disasters strike, about 10\% of counties in a region are affected, based on FEMA emergency declarations.  
To assess the responses to a shock hitting 10\% of counties, as an example, the reported IRFs should be divided by 10. Thus, for instance, a shock of this size occurring in the South can reduce growth by 0.09\%-0.18\% in other regions, approximately.

Further results and robustness checks are presented in Section \ref{sec:moreres}. Moreover, in the Appendix we also replicate the analysis for different categories of weather-related disasters.
\FloatBarrier
\subsection{The role of economic linkages and spatial proximity}\label{sec:spatial}

\noindent The GVAR model allows us to estimate spillover effects taking into account the economic interdependence between states. In this section, we assess the importance of explicitly considering economic linkages through interstate trade flows. To this aim, we re-estimate the GVAR model using alternative weight matrices $\mathbf{W}_i$ based on spatial adjacency.

 \begin{figure}[t]
\caption{IRFs using spatial adjacency matrix, 12-month horizon}
\centering
\includegraphics[width=\textwidth]{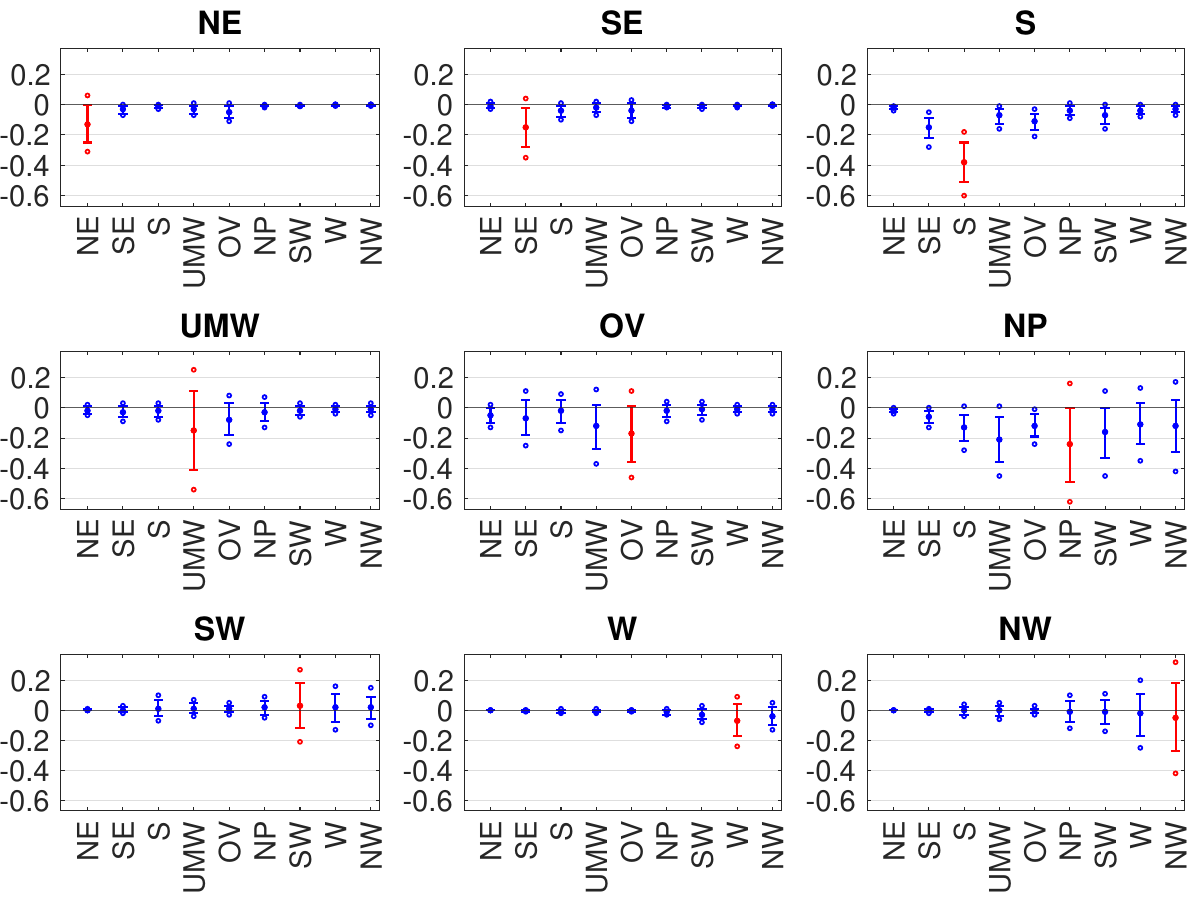}
    \caption*{\scriptsize\textit{Notes:} See notes to Figure \ref{fig:figirf_all_impact}.}
    \label{fig:irf_all_adjacency}
\end{figure}

Figure \ref{fig:irf_all_adjacency} reports the IRFs of MECI calculated by considering geographical proximity. 
Cross-region effects of weather shocks are almost entirely wiped out, and intra-regional effects are substantially reduced, relative to the baseline results in Figure \ref{fig:figirf_all}.\footnote{Figures \ref{fig:figirf_all_impact}-\ref{fig:irf_all_adjacency} are reported in tabular form in the online Appendix.}

\subsection{The role of parameter heterogeneity and disaggregation}\label{sec:hetero}
\noindent\textit{A spatial dynamic panel model.} In the GVAR specification, the parameters are heterogeneous across states. To highlight the importance of such heterogeneity, we compare the baseline GVAR results with the estimates of a spatial dynamic model (i.e., a model with both space and time lags) in which all states share the same parameters:
 \begin{equation}\label{eq:sdm}
     y_{it} = p \mathbf{\widetilde{W}}_i \mathbf{y}_t + \gamma y_{i,t-1} + \rho  \mathbf{\widetilde{W}}_i \mathbf{y}_{t-1} + \beta s_{it} + c_{i} + e_{it}
 \end{equation}
 where $\mathbf{\widetilde{W}}_i$ denotes an $N$-dimensional row vector of trade weights, $p, \gamma, \rho, \beta$ are homogeneous scalar parameters, $c_i$ denotes state fixed effects and $e_{it}$ is an error term. We use a tilde to distinguish the weight vector employed here from the $2 \times N$ weight matrix used in \eqref{eq:GVAR_weights}, but the bilateral weights are exactly the same in the two models (i.e., $ y_{it}^\ast  = \mathbf{\widetilde{W}}_i \mathbf{y}_t$).  \citet{YuLee2008} have developed quasi-maximum likelihood estimators for spatial dynamic panel models of the form \eqref{eq:sdm}. In panel (\textit{a}) of Table \ref{tab:sdm_est}, we report bias-corrected estimates of model \eqref{eq:sdm}, using the \citet{YuLee2008} method. The coefficient on weather shocks is around $-0.074$ and statistically different from zero.

\begin{table}[t]
 \caption{The role of parameter heterogeneity and disaggregation}
 \centering
\begin{tabularx}{\textwidth}{XXX}
\hline
\multicolumn{3}{l}{(\textit{a}) {A spatial dynamic model}}\\\hline
variable & coefficient & std. error \\
\hline
$\mathbf{\widetilde{W}}_i \mathbf{y}_t$ & 0.7109***    &  0.0089  \\
$y_{i,t-1}$ & 0.1579*** & 0.0074 \\
$\mathbf{\widetilde{W}}_i \mathbf{y}_{t-1}$ & $-0.0392$***  & 0.0121  \\
$s_{it}$ & $-0.0737$*** & 0.0099 \\
\hline 
\multicolumn{3}{l}{(\textit{b}) An aggregate U.S. ARDL model}\\\hline
variable & coefficient & std. error\\
\hline 
$y_{t-1}$ & 0.6037*** 
               & 0.0529 \\
$y_{t-2}$ & 0.0811 
               & 0.0530 \\
$s_t$      & $-0.2255$
               & 0.1572\\
$const$          & 0.0083 
               & 0.0118\\
\hline
\end{tabularx}
\begin{tabularx}{\textwidth}{XXXXX}
\multicolumn{5}{l}{(\textit{c}) Heterogeneous coefficients of $s_{it}$ in the GVAR}\\\hline
parameter & average & std. dev. & min  & max \\\hline
$\theta_i$ & $-0.067$ & $0.11$ & $-0.61$ & $0.065$ \\
\hline
\end{tabularx}
\caption*{\scriptsize\textit{Notes:} 
Panel (a) reports estimates of a spatial dynamic panel model. The estimates of an autoregressive distributed lag (ARDL) model for the aggregate U.S. economy are shown in panel (b). Panel (c) displays the average point estimate of coefficient $\theta_i$ (i.e., the impact effect of weather shocks) across the state-specific ARX* models \eqref{eq:ARX} composing the GVAR, along with the standard deviation of the estimates across states, and the minimum and maximum estimates. In panels (a) and (b), $^{\ast\ast\ast}$ indicates 1\% significance, the other coefficients in panel (b) are not significant at the 10\% level.
}
\label{tab:sdm_est}
\end{table}
\FloatBarrier
 
\bigskip

\noindent\textit{A country-wide regression for the U.S.} Next, we consider a model that does not feature geographical disaggregation.
Specifically, we estimate an autoregressive distributed lag (ARDL) model for the aggregate U.S. economy at the monthly frequency. As shown in  panel (\textit{b}) of Table \ref{tab:sdm_est}, the effect of weather shocks is negative but statistically indistinguishable from zero. We conclude that disaggregation matters, when it comes to estimating the economic impact of weather events.

\bigskip

Finally, for comparison, panel (\textit{c}) of Table \ref{tab:sdm_est} reports the average point estimate of the on-impact effect of weather shocks, $\theta_i$, across the state-specific ARX* models \eqref{eq:ARX} composing the GVAR, along with the standard deviation of the estimates across states, as well as the minimum and maximum estimates.
The average estimate of $\theta_i$ in the GVAR model ($-0.067$) is similar to the coefficient on weather shocks in the spatial dynamic model ($-0.074$). 
However, the dynamic panel model fails to capture the heterogeneity of the response to weather shocks across states.
As highlighted in panel (\textit{c}), the estimated effects of weather shocks vary substantially across states in the GVAR model: 
the point estimates of coefficient $\theta_i$ have a standard deviation of $0.11$ across states, with values in the range from $-0.61$ to $0.065$ (for Louisiana and Maine, respectively).

\subsection{Indirect effects of weather shocks on local activity}
\noindent To quantify indirect effects of weather shocks on local activity for each state, we compare the IRFs derived from the GVAR model with alternative IRFs derived using state-specific ARDL (ARX*) models and assuming that changes in state-specific variables do not affect other states. 
The difference between the IRFs in the two cases captures ``second-round effects'': 
a weather shock in state $i$ causes economic adjustment in the other states, which in turn feeds back into state $i$.

\begin{figure}[t]
    \centering
\caption{Second-round effects of weather shocks}
        \includegraphics[width=1\linewidth]{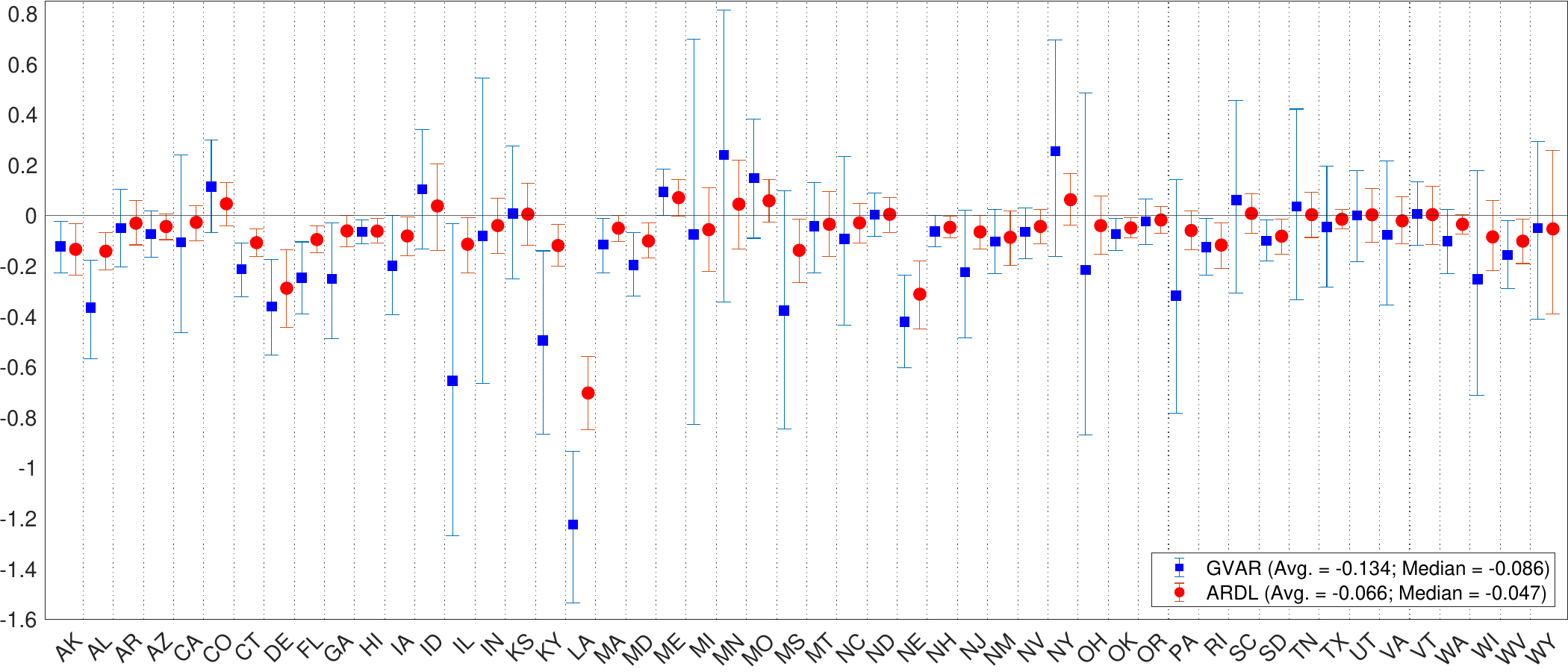}
\caption*{\scriptsize\textit{Notes:} For each state, the figure reports the bootstrap mean estimates of 12-month-ahead responses of state-level MECI to a simulated weather shock in the same state, provided by the U.S. GVAR model (blue squares) and by state-specific ARDL ($ARX^\ast$) models in which the rest of the U.S. economy is exogenous (red circles). The bars indicate the 10th and 90th percentiles from the bootstrap distribution.}
\label{fig:second_round}
\end{figure}

Figure \ref{fig:second_round} isolates second-round effects by comparing GVAR and ARDL estimates of the response of MECI to a weather shock after one year. In absolute terms, these effects are especially strong in Louisiana (LA) and Illinois (IL). The results for the former are largely driven by the devastating impact of Hurricane Katrina in 2005, while those for the latter are especially associated with extreme winter weather events, such as the Blizzard of 1999. 

More generally, we note that, as expected, GVAR estimates have wider confidence intervals but also tend to be greater in absolute value, hence reinforcing the conclusion that spillovers do matter in assessing the impact of weather shocks. As a back-of-the-envelope calculation, the legend of the figure indicates that, in absolute terms, both the average and median responses from the GVAR are larger than those from the ARDL models.

\FloatBarrier

\section{Further results and robustness checks}\label{sec:moreres}

\noindent In this section, we present a number of additional results and robustness checks. For the sake of brevity, the tables and figures underlying these results are presented in the online Appendix, to which the reader is referred.

\subsection{Alternative proxies of weather related disasters}\label{sec:aci}

\noindent In this section, we consider two alternative proxies of weather shocks: one based on ex-ante impact estimates generated using the CLIMADA modelling framework, and the other on the Actuaries Climate Index, which provides a composite measure of climate extremes -- including temperature anomalies, heavy precipitation, prolonged dry spells, high winds, and sea level changes -- relevant for assessing economic vulnerability.

\bigskip

\begin{figure}[!t]
		\caption{Gridded impact analysis in CLIMADA: an example with Florida}
		\centering
		\includegraphics[width=.36\textwidth]{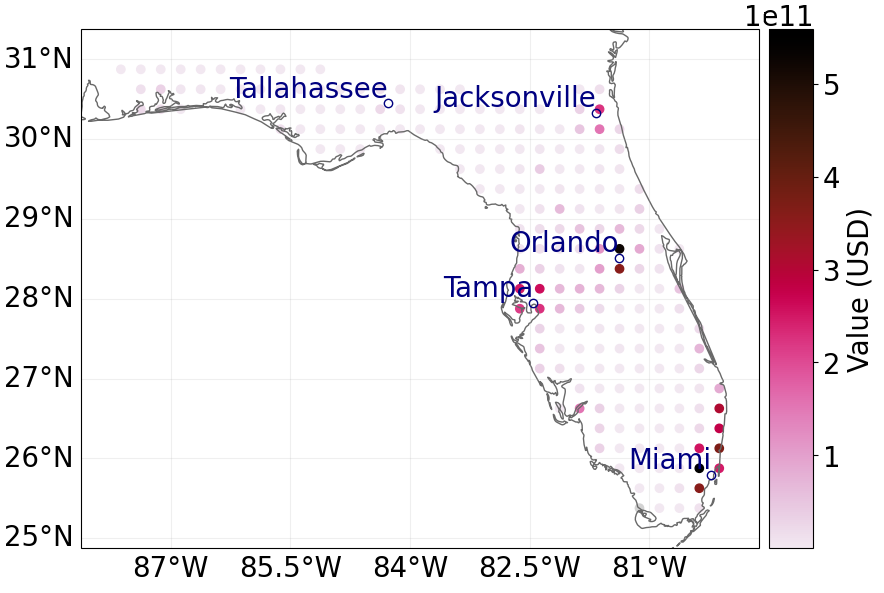}%
       	\includegraphics[width=.34\textwidth]{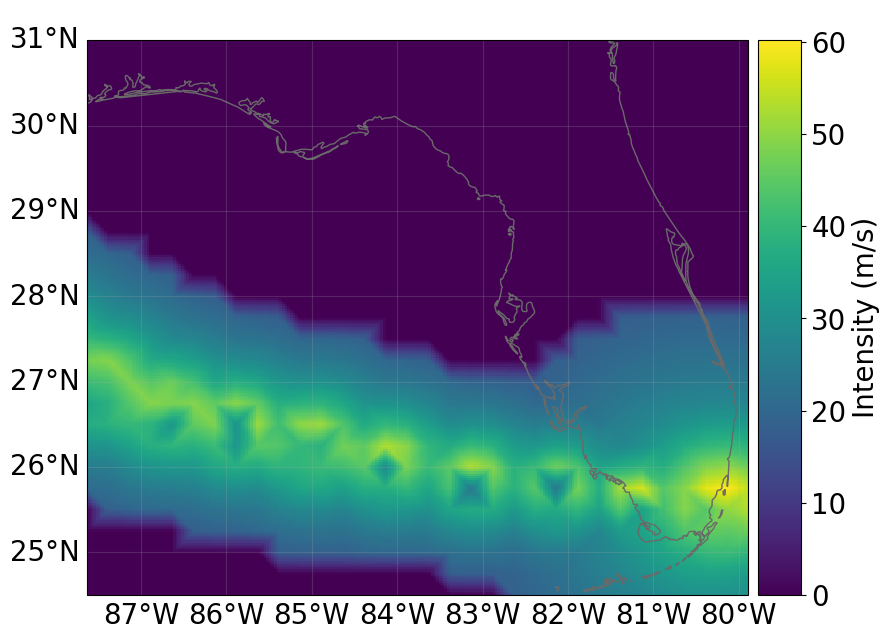}%
        \includegraphics[width=.31\textwidth]{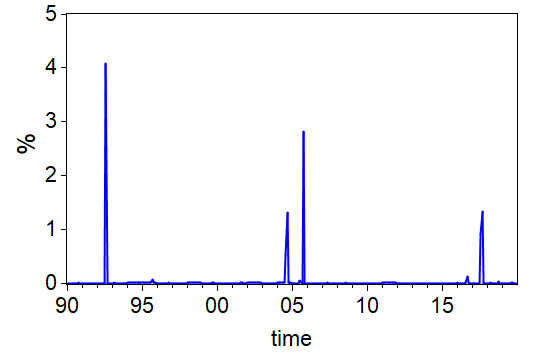}

		\subcaption*{\scriptsize\textit{Notes:} The figure on the left shows gridded economic exposure, based on night-time lighting and population density. Each dot represents a centroid, and the color scale indicates the level of exposure. For illustrative purposes, the absolute exposure scale is set using total household wealth, as provided by CLIMADA based on Credit Suisse data. The central figure shows the maximum intensity of Hurricane Andrew, measured by wind speed in meters per second (m/s), in Florida in August 1992, based on IBTrACS data. The figure on the right shows the CLIMADA-based indicator of weather shocks for Florida, in \% of total state assets, from 1990m1 to 2019m12. This indicator is constructed using high-resolution data on tropical cyclones, other storms and river floods.}
		\label{fig:litpop_florida}
	\end{figure}

\noindent \textit{CLIMADA}. The first alternative proxy is based on a measure of the ex-ante estimated impact of weather events, based on the physical features of the events and the economic exposure of the affected areas. To this end, we make use of CLIMADA (CLIMate ADAptation), a free and open-source software framework\footnote{For more information on the current version, see the project website at \url{https://climada.ethz.ch/}.}, specifically designed to quantify the impacts of natural hazards and to assess climate risk \citep{aznar2019climada}. CLIMADA allows making projections of the damages from natural events at high geographical resolutions, by integrating data on hazards, data on exposures, and measures of vulnerability in the form of impact functions mapping the physical severity of hazards into economic losses.\footnote{Specifically, an \textit{impact function} determines the percentage of damage to an exposed asset as a function of hazard intensity. In the modelling community, this is often referred to as a \textit{vulnerability curve}.} Further details on the underlying data and CLIMADA modules are provided in the Appendix.\newline
\indent For each U.S. state, we use CLIMADA to generate a time series of projected losses caused by several types of weather events, closely matching those considered in the FEMA classification.
Specifically, we calculate losses caused by tropical cyclones, other types of storms, river floods and wildfires. We first calculate the daily losses caused by each single event using granular geographic information (i.e., gridded data at a high resolution) on the physical characteristics of events and on the economic exposures. Then, we aggregate the estimated losses at the state level, at the monthly frequency, and across different events, to have a time series of losses that can be used in the GVAR in place of the FEMA-based proxy.\newline
\indent Figure \ref{fig:litpop_florida} provides an example for a representative state, Florida. The leftmost figure plots the gridded economic exposures based on nightlights and population, the central figure plots the maximum intensity at each location of a major event, namely Hurricane Andrew of August 1992 (the intensity is measured by wind speed in m/s), and the figure on the right plots the time series of projected losses, reported as percentages of total state assets. Of course, the largest exposures are associated with main urban centers: Miami, Orlando, Tampa, Jacksonville. The peaks in the projected impact series are associated with major hurricanes, like Hurricane Andrew (August 1992) or Hurricane Wilma (October 2005).

Using this approach, we find that weather shocks that generate significant effects are those hitting the South and Southeast regions, while shocks in other regions have generally non-significant impacts. While the results for the South and Southeast regions are totally in line with our baseline results based on the FEMA proxy, the main differences (in particular, the insignificant impact of shocks in the Northeast) appear at least in part due to the different way of dealing with winter weather events, which in CLIMADA we capture only to the extent that they are associated with storm activity.

\bigskip

\noindent \textit{Actuaries Climate Index (ACI)}. We compare our results also with those obtained using the Actuaries Climate Index (ACI) as an alternative proxy of weather related disasters \citep[see, e.g.,][]{KimMatthesPhan2011}. The ACI measures the frequency of extreme weather events and the extent of sea level change on a monthly basis, condensing information related to six variables: high and low temperatures, heavy precipitation, consecutive dry days, high winds, and sea level.

\begin{figure}
    \centering
        \caption{Weather Related Disasters and the Actuaries Climate Index (ACI)}
    \includegraphics[width=.8\linewidth]{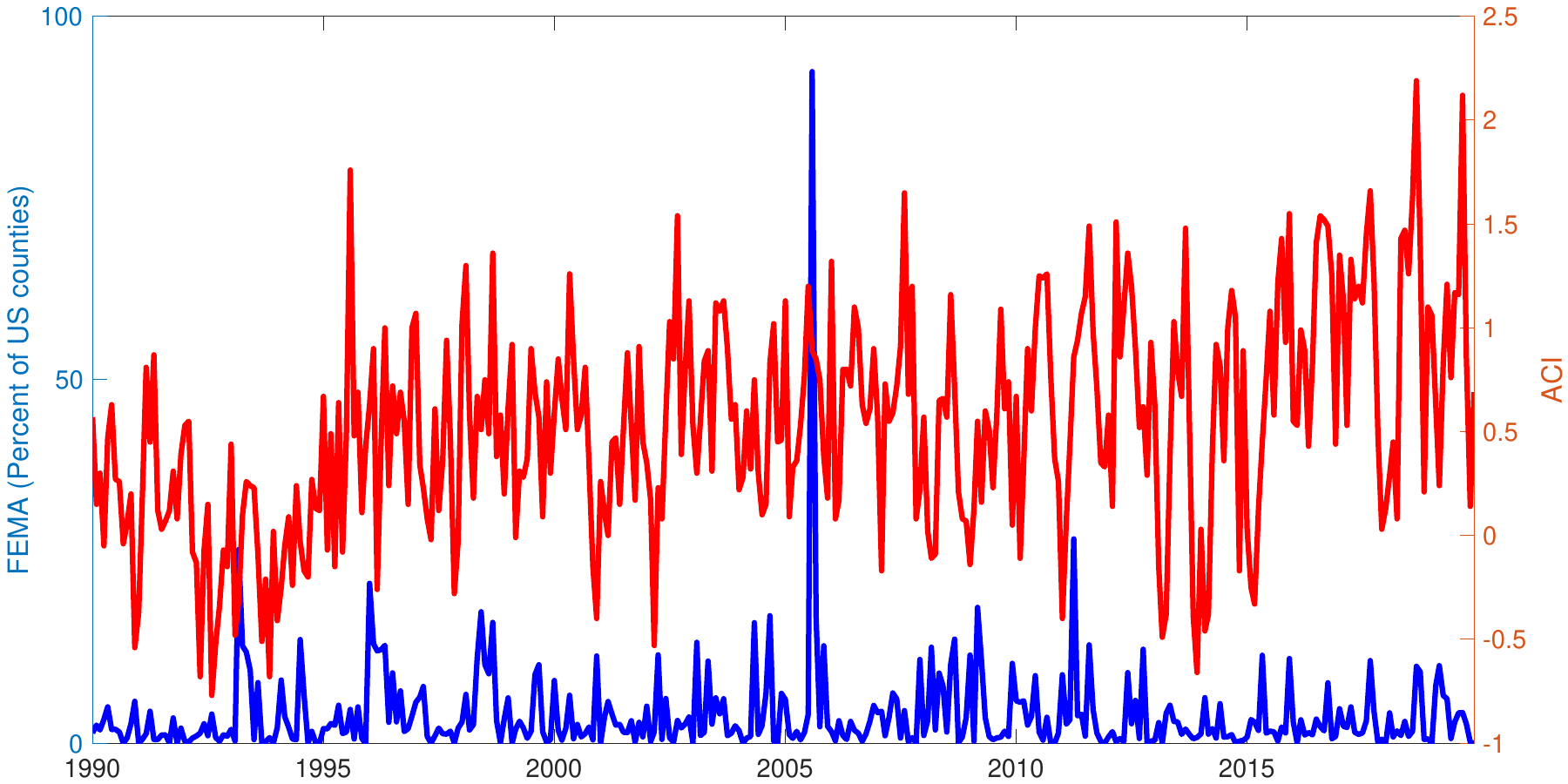}
    \label{fig:FigACI}
    \caption*{\scriptsize{\textit{Notes}: The figure plots the (unsmoothed) ACI for the U.S., along with our FEMA-based indicator. Data on ACI are sourced from \url{https://actuariesclimateindex.org}}}
\end{figure}

Figure \ref{fig:FigACI} plots both the ACI and our proxy for weather-related disasters for the U.S. Visual inspection of this plot reveals that the two time series exhibit very different patterns. In particular, the ACI shows a mean shift at the beginning of 1995; this is consistent with the empirical evidence in \citet{KimMatthesPhan2011}, which highlights that both the mean and variability of the ACI are on average higher in recent years than in the past. Such behavior of the series suggests that the frequency of severe weather conditions, as captured by the ACI, has gradually increased over time. In contrast, our proxy for weather disasters does not capture the variation in the frequency of such weather conditions, but is more like a dummy for the occurrence of extreme events.
The ACI is not available at the state level, but for seven regions: Alaska (ALA), Central East Atlantic (CEA), Central West Pacific (CWP), Midwest (MID), Southeast Atlantic (SEA), Southern Plains (SPL), Southwest Pacific (SWP).

After properly aggregating state-level data at ACI regional level, we estimate a GVAR model using alternatively our measure of weather-related disasters and the ACI index as exogenous variables. In the case of our proxy, the estimated effects of weather shocks remain negative and generally significant at the 10\% (one-sided), with the only exceptions of shocks hitting the SPL and SWP regions. When we use ACI, significance at the 10\% is achieved only for shocks hitting the CWP region. Overall, the results suggest that focusing on emergency declarations triggered by weather-related disasters may provide sharper identification of disruptive weather shocks, compared to indicators of abnormal climate conditions.

\subsection{On the transmission channels of weather shocks}\label{sec:channels}

\noindent In this section, we discuss the economic channels through which extreme weather events affect local economic activity and generate cross-border spillover effects.

\bigskip

\noindent\textit{MECI components.} 
We start by re-estimating the GVAR model using, alternatively, the six sub-indices of the aggregate MECI indicators: (\textit{i}) mobility, (\textit{ii}) labor market, (\textit{iii}) real activity, (\textit{iv}) expectations, (\textit{v}) households, (\textit{vi}) financials. 
The complete set of results as well as details on the MECI sub-indices are reported in the Appendix.

The component of MECI that is associated with the strongest effects of weather shocks (both local and cross-border) and with high statistical significance in most regions is the ``labor market'' component. Another component that shows statistical significance in several regions is ``mobility'', although in this case the magnitude of the impact and, more generally, the contribution of the component to total MECI are quite small. The responses of the ``real activity'', ``households'', and ``expectations'' components are non-negligible only in the case of shocks hitting the South, while the effects on the ``financials" component are invariably non-significant.

\bigskip

\noindent\textit{Labor-market indicators.} Next, to characterize in more detail the predominant labor-market channel, we study the responses of two specific indicators of labor activities: the number of hours worked in manufacturing and the growth rate of total employment. Both indicators generally drop after a weather-related disaster, both locally and in regions not directly hit. In the Appendix we also consider the impact of different types of weather events on employment. Winter weather, storms and floods have a significant negative impact in several regions, while tropical storms affect the Southeast in particular. Droughts and fires have no significant impact.

\bigskip

\noindent\textit{Weather shocks as supply-side shocks.} Overall, the previous results, indicating a substantial impact of extreme weather on labor market conditions and mobility and, at the same time, a limited impact on typically demand-related components like ``households'' and ``expectations'', suggest a central role of supply-side disruptions.
To test that weather shocks are mainly supply-side shocks, we look at movements in consumer prices. Unfortunately, monthly inflation data are not available at the state level. However, the U.S. Bureau of Labor Statistics (BLS) publishes monthly data on 12-month inflation rates for four macro-regions from September 2004. The four regions are Northeast, Midwest, South, and West, and cover the entire U.S. territory.

A descriptive analysis of the data suggests that, following major weather-related disasters, inflation rates tend to increase across the entire U.S., regardless of where the event occurred. For instance, in August-September 2005, following Hurricane Katrina, the 12-month inflation rate increased by 2.4 percentage points in the Northeast, 2 in the Midwest, 1.8 in the South, and 1.1 in the West. Analogously, in April-May 2011, following the so-called Super Outbreak, the inflation rate increased by 0.7 percentage points in the Northeast, 1.1 in the Midwest, 1.2 in the South, and 0.6 in the West.

To test more formally that all regional inflation rates tend to increase in response to weather shocks, we estimate a VARX model for the four regional inflation rates, including our FEMA-based weather shock variable for the aggregate U.S. territory as a common exogenous variable.\footnote{To control for the effects of aggregate economic activity and monetary policy on inflation, we also include the (12-month) growth rate of U.S.-wide industrial production and the Fed funds rate as two additional common exogenous variables. To account for potential price stickiness, we include both contemporaneous and lagged values of the exogenous variables. The lag order of the VARX is set to one, as suggested by the BIC.} Thus, we consider the responses of regional prices to extreme weather events, regardless of where the events occurred.

The results confirm generalized price increases across regions, especially significant in the Midwest and in the South. This might suggest that the high degree of market integration across the U.S., as reflected in the strong co-movement of regional consumer price inflation rates, may act as a powerful mechanism underlying the cross-state transmission of shocks, thus helping explain the significant spillover effects found in our main results.

To sum up, our analysis so far suggests two main conclusions regarding the channels of transmission of weather-related disasters: (\textit{i}) weather shocks mainly affect the local economy through supply and labor market disruptions; (\textit{ii}) as a result of market integration, supply contractions appear to spread across the U.S., even if not all regions are directly hit by extreme events. However, further evidence will be needed to corroborate this interpretation of the results and to study supply-side effects. While we mostly leave this to future research, below we further explore supply-side contagion as a source of spillovers.

\bigskip

\noindent \textit{Inspecting the mechanisms behind spillovers.} Lastly, since our main results reveal that geographic spillovers from severe weather events are strongly tied to economic interconnections between states, we now seek to further investigate the underlying causes and transmission mechanisms of these spillover effects.\\
\indent The literature highlights the critical role of supply-chain (input-output) relationships in the transmission of natural disaster shocks. \citet{BarrotSauvagnat2016} show that disasters disrupt production networks in the U.S. by affecting supplier firms, thereby causing significant indirect losses to downstream customer firms -- even in unaffected areas -- due to costly input substitution. Similarly, \citet{hallegatte2008adaptive,hallegatte2014modeling} provides evidence of substantial national economic losses from disasters like Hurricane Katrina via sectoral linkages. International spillovers are also observed, as shown by \citet{feng2024we}, \citet{ForslidSanctuary2023}, \citet{Kashiwagi2021}, and \citet{BoehmFlaaenPandalaiNayar2019}, who find that disasters impact countries indirectly through global supply chains. Building on this literature, we present two quantitative assessments -- each employing a distinct methodological approach -- of the origins and magnitude of spillover effects. Methodological details and full results are provided in the Appendix.\newline
\indent In the first exercise, we return to our GVAR model to investigate whether economic spillovers are more strongly linked to trade in intermediate goods rather than final goods, as suggested by earlier studies. Although the available data do not explicitly classify interstate shipments by final or intermediate use, we rely on the sectoral information provided by the U.S. Commodity Flow Survey to make this distinction. By combining these data with input-output tables from the U.S. Bureau of Economic Analysis, we classify sectors based on whether the majority of their output is used as intermediate inputs or final goods.\newline
\indent Using this classification, we build two separate trade-weight matrices, $\mathbf{W}$: one reflecting trade in intermediate-good-intensive sectors and the other for final-good-intensive sectors. We then re-estimate the GVAR model using either matrix separately. The results indicate that spillovers are stronger when focusing on trade in intermediate goods, thus reinforcing the idea that input linkages across regions and sectors play a key role in amplifying the effects of economic shocks.\newline
\indent The second exercise explores the indirect economic consequences of natural disasters by focusing on how disruptions in one part of the economy can ripple across sectors. Using the CLIMADA software, the analysis focuses on tropical cyclones in the United States from 1990 to 2019. We estimate both the immediate (direct) damages in affected areas and the broader (indirect) nationwide impacts which arise from input-output relationships between economic sectors.\newline
\indent As shown in Figure A13 in the Appendix, indirect losses are substantial, nearly half the size of direct damages. For instance, in 2005, the year of Hurricane Katrina, direct losses from tropical cyclones are estimated at around 0.75\% of GDP, while indirect losses add another 0.3\%. Although this is a static analysis that does not account for longer-term economic adjustments, it still underscores the significant role of supply-chain linkages in amplifying the economic costs of natural disasters.

\subsection{The possible endogeneity of the weather shock indicator}\label{sec:endog}

\noindent The political influences on emergency declarations and the quantification of relief funding have been addressed in different studies, e.g., \cite{GarrettSobel2003}. Our indicator of weather shocks, being based on emergency declarations, might contain information not strictly related to natural disasters, with possible concerns about endogeneity.\footnote{We thank an anonymous referee for raising this interesting issue. \cite{FelbermayrGroschl2014} propose a dataset (denoted GeoMet) containing measures of the physical strength for a set of natural disasters. Such indicators, not based on monetary damages, solve the potential endogeneity problem, as well as, in studies relating to very different countries, the issue of higher damages being generally associated with richer countries. On the other hand, these indicators, based on different measurement scales, might give rise to aggregation issues, if the aim is to define a unique indicator of natural disasters. Moreover, particular types of natural disasters, like wildfires, are very difficult to measure with a single indicator.} 

To deal with this potential identification issue, we propose a robustness check based on a two-stage approach, where in the first stage our indicator of natural disasters is regressed, state by state, on the number of deaths and a dummy variable denoting whether the disaster is included in the U.S. Billion-Dollar Weather \& Climate Disaster database.\footnote{As the FEMA and the Billion-Dollar Weather \& Climate Disaster databases are not perfectly synchronized, especially for events that occurred over long time spans (e.g., droughts, wildfires), or occurred across two distinct months, we preliminarily performed a deep data cleaning to match events in the two databases. All modifications, however, have been implemented when an appropriate documentation was found.} 

The results of the first stage regressions confirm that this exogenous external information significantly helps to capture the main features of the weather shock indicator. Even more important, in the second stage, where the weather shock indicator is replaced by the endogenous-free fitted values of the first stage, almost all the results described in Section \ref{sec:mainresults} are confirmed, both in terms of the impact on the local economic activity measure and of the spillover effects.

To further address potential endogeneity, we also introduce an alternative instrumental variable based on extreme weather events. Our earlier approach relied on the number of deaths caused by such events, using it as a proxy for severity, while deliberately avoiding monetary measures of impact. We now refine this strategy by considering an instrument that is entirely independent of any type of impact -- monetary or non-monetary, observed or expected. Rather than reflecting consequences, this instrument is grounded solely in the physical attributes of natural events and the ex-ante exposure of the affected areas.\newline
\indent To construct this alternative instrumental variable, we follow \citet{bilal2023anticipating} and build a storm indicator that reflects the physical severity of weather phenomena (specifically, maximum precipitation and wind speed—across U.S. states). The construction involves removing state and seasonal fixed effects from the raw weather variables, and isolating extreme realizations via residuals exceeding the 95th percentile. The final indicator, $D^{storm}_{it}$, captures whether a state experiences either unusually high precipitation or wind speed in a given month. To incorporate the economic relevance of these events, the storm variable is weighted using gridded economic activity data.\footnote{The underlying climate data come from the Weighted Climate Dataset (WCD) developed by \citet{lampertidata}, which integrates socio-economic and climate metrics. Further details are provided in the Appendix.} This storm indicator is then used as an instrument for our FEMA-based disaster variable in a two-stage least squares estimation within the GVAR framework. The results largely confirm our baseline findings, with significantly negative effects in the Southeast, South, and Ohio Valley regions, although effects in the Northeast are no longer statistically significant in this case.

\subsection{Robustness checks}

\noindent\textit{Nonlinearities.} We rely on two different approaches to assess potential nonlinearities in the relationships between weather shocks and economic activity. First, we re-estimate the U.S. GVAR adding the square of the weather shock variable as an additional regressor. Second, we allow for the possibility that the negative impact of adverse weather events on a state's economy may be amplified if the state has already experienced a weather shock in the recent past. To this end, we include an interaction term between our weather shock variable $s_{it}$ and a dummy variable indicating whether the state experienced a severe weather shock in the last year. In both cases, results remain qualitatively similar to those in Section \ref{sec:mainresults}.

\bigskip

\noindent\textit{Including state-level control variables.} The inclusion of RoUS averages in state-specific ARX* models allows us to account for common factors affecting the aggregate U.S. economy. We now check whether our results are robust to the inclusion of state-specific variables that may affect state-level economic activity. We thus re-estimate the GVAR model including two additional regressors in each ARX* model: the government component of Gross State Product and state exports to the rest of the world. As in our baseline results, the effects of weather shocks are negative and at least one-sided 10\% significant for shocks hitting the Northeast, Southeast, South and Northern Rockies and Plains regions. In addition, negative and significant results are now obtained also for the West and Northwest regions.

\bigskip

\noindent \textit{Alternative county weights.} Our baseline weighting of weather events, based on the percentage of counties hit in a state, is intended to capture the intensity of disasters rather than the importance of individual counties. As a robustness check, however, we consider an alternative weighting scheme reflecting the different population levels of affected counties. Specifically, for each state, we scale the weather event dummy using the population of counties hit as a share of the state's total population.  
Once again, full results are reported in the Appendix.
The effects of weather shocks remain significantly negative for events hitting the Northeast and South regions, and close to significance for events hitting the Southeast. Thus, the results are broadly in line but weaker than the baseline. This appears consistent with the fact that productive establishments (such as factories), whose activities are disrupted by extreme weather events, are often located in areas that are not densely populated.

\bigskip

\noindent\textit{Further robustness checks.} We report additional robustness checks in the Appendix. These include re-estimating the model over a shorter sample that ends before the Global Financial Crisis of 2007–2009, assessing the potential endogeneity of the trade weights, and examining their time constancy. We also test the robustness of our results to alternative specifications of the trade-weight matrices.

\section{Conclusions and policy implications}\label{sec:conclusions}
 
\noindent We have investigated the short-run effects of weather shocks on economic activity across U.S. states, with a particular focus on cross-border spillover effects. Using a GVAR framework, we have shown that severe weather shocks generally cause drops in state-level economic activity, with negative effects propagating to other regions through economic linkages between states. We estimate that an extreme weather event involving one climate region of the U.S. has the potential to reduce regional yearly growth rates by up to 1.8 percentage points, approximately. Because of indirect (second-round) effects, network relationships between states amplify the total impact that weather shocks have on the states where they occur.

Our results provide several implications from a policy perspective. Moreover, such implications concern both federal and local governments. On the one hand, the short-run negative effects on the economy of the region directly affected by an adverse weather shock suggest that local policymakers should plan in advance about the possible consequences of sudden increases in spending needs and, at the same time, declines in the revenues deriving from taxation. On the other hand, our results about economic spillovers reveal that all the other states (not just those geographically close to the area affected by the natural disaster) are not immune from serious adverse economic consequences. This evidence suggests the need for coordination of local policies to shield regional economies from severe weather shocks, but also for a direct and proactive intervention of the federal government to supervise and coordinate local policies, providing strategic actions for dealing with extreme natural events which may become more and more frequent in the future. 

\vspace{1cm}
\bibliography{BBM}

\end{document}